\documentclass[a4paper,12pt]{article}

\usepackage{amsmath,amsfonts,amssymb,epsfig}

\numberwithin{equation}{section}

\usepackage{cancel}
\usepackage{cite}
\usepackage{pstricks}
\usepackage{axodraw4j}
\usepackage{pmat}

\usepackage[T1]{fontenc}
\usepackage[latin9]{inputenc}

\usepackage{a4wide}

\def\psqr#1#2{{\vcenter{\vbox{\hrule height.#2pt
        \hbox{\vrule width.#2pt height#1pt \kern#1pt
        \vrule width.#2pt}
        \hrule height.#2pt \hrule height.#2pt
        \hbox{\vrule width.#2pt height#1pt \kern#1pt
        \vrule width.#2pt}
        \hrule height.#2pt}}}}
\def\sqr#1#2{{\vcenter{\vbox{\hrule height.#2pt
        \hbox{\vrule width.#2pt height#1pt \kern#1pt
        \vrule width.#2pt}
        \hrule height.#2pt}}}}

\newcommand{\fund}{\mathchoice\sqr65\sqr65\sqr{2.1}3\sqr{1.5}3}
\newcommand{\afund}{\ov{\mathchoice\sqr65\sqr65\sqr{2.1}3\sqr{1.5}3}}

\newcommand{\SU}[1]{\mathrm{SU}(#1)}
\newcommand{\U}[1]{\mathrm{U}(#1)}

\newcommand{\sunc}{\mathrm{SU}(\nc)}
\newcommand{\sunnc}{\mathrm{SU}(\nnc)}
\newcommand{\sunf}{\mathrm{SU}(\nf)}
\newcommand{\nc}{N}
\newcommand{\nnc}{n}
\newcommand{\nf}{{{N}_f}}
\def\bra{\langle}
\def\ket{\rangle}

\def\tr{\mathrm{tr}}

\def\beq{\begin{equation}}
\def\eeq{\end{equation}}
\def\bal{\begin{align}}
\def\eal{\end{align}}
\def\nn{\nonumber}

\def\2b2[#1,#2][#3,#4]{\left( \begin{array}{cc} #1 & #2 \\ #3 & #4 \end{array} \right)}
\def\3b3[#1,#2,#3][#4,#5,#6][#7,#8,#9]{\left( \begin{array}{ccc} #1 & #2 #3 \\ #4 & #5 & #6\\#7&#8&#9\end{array} \right)}

\newcommand{\C}[1]{\mathcal{#1}}

\def\ov{\overline}

\author{\vspace{-2.2cm}
Steven~Abel$^{1,2}$\footnote{s.a.abel@durham.ac.uk}~
 and Mark~Goodsell$^3$\footnote{mark.goodsell@desy.de}}
\date{}
\title{{}
\vspace{-4cm}
\hfill{\tiny{DESY 11-010}}\\
\vspace{-0.5cm}
\hfill{\tiny{IPPP/11/03; DCPT/11/06}}\\
\vspace{-0.5cm}
\hfill{\tiny{CERN-PH-TH/2011-009}}\\[5.8cm]}

\begin{document}

\maketitle
\begin{center}
{\vspace{-5.5cm}
{\includegraphics[width=0.65\textwidth]{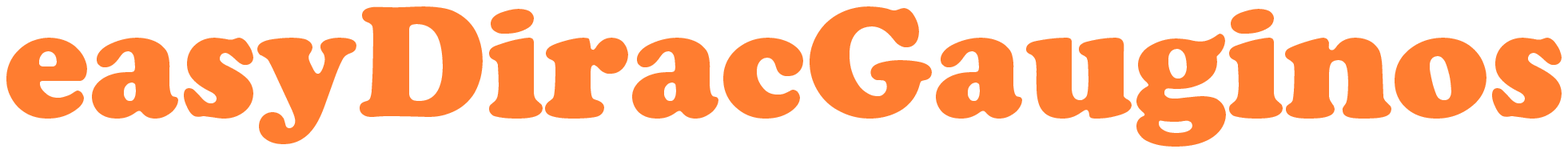}\\[-8.0cm]    
\vspace{-3cm}\emph{$^1$Institute for Particle Physics Phenomenology, Durham University,
UK\\
$^2$Theory Division, CERN, 1211 Geneva 23, Switzerland\\
$^3$Deutsches  Elektronen-Synchrotron, DESY, Notkestra\ss e 85, 22607  Hamburg, Germany}}
\vspace{1.5cm}}
\end{center}

\abstract{ \noindent A simple and natural model is presented that gives Dirac gauginos. The configuration is 
related to ``deconstructed gaugino mediation''. A high energy completion is provided based on existing
ISS-like models of deconstructed gaugino mediation. This provides a complete picture of Dirac gauginos that includes the necessary extra adjoint fermions (generated as magnetic quarks of the ISS theory) and supersymmetry breaking (via the ISS mechanism). Moreover the screening of the scalar masses means that they can similar to or less than the gaugino masses, 
even though the supersymmetry breaking is driven by $F$-terms.}

\newpage

\section{Introduction}

$R$-symmetry is an important aspect of supersymmetry breaking because
it is directly related to the existence or otherwise of global supersymmetric
minima in generic theories \cite{Nelson:1993nf}, and yet Majorana gaugino masses are bound to break it. 
Considerable effort has therefore been devoted to 
the problem of how to generate acceptably large Majorana gaugino masses whilst 
maintaining enough $R$-symmetry to protect supersymmetry breaking. 

An interesting observation
that followed on from the work of \cite{ISS} (ISS) was that strong dynamics
can produce an\emph{ emergent} $R$-symmetry\cite{Kitano:2006xg,MN}.
The authors of Ref.\cite{MN} in particular used the fact that in theories such as ISS, operators that are irrelevant in the
ultra-violet (UV) can become marginal in the infra-red (IR). Dimensional
arguments then indicate that the couplings of such operators can be
highly suppressed in the IR, and this can in turn lead to approximate
$R$-symmetries at low energies, which are preserved, but for these
small effective couplings. 

Such emergent $R$-symmetry can indeed help with Majorana masses. But given the close link between strong dynamics,
$R$-symmetry and supersymmetry breaking, it seems interesting to ask
if the dynamics of strongly coupled 
supersymmetric QCD (SQCD) can instead produce configurations of gauge mediation
that have {\em Dirac} gaugino masses. These are by contrast able to respect the 
all-important $R$-symmetry required by supersymmetry breaking, and so can be advantageous
from the metastability point of view. Their numerous other advantages 
have been documented in a wide ranging program of work~\cite{Fayet:1978qc, Polchinski:1982an, Hall:1990hq, Fox:2002bu, Nelson:2002ca,Antoniadis:2005em,Antoniadis:2006uj,Hsieh:2007wq, Amigo:2008rc, Benakli:2008pg, Belanger:2009wf,Benakli:2009mk, Chun:2009zx, Benakli:2010gi,Carpenter:2010as,Kribs:2010md}. This is the subject of our paper. 

One problem that has in the past hampered Dirac gauginos is that, when there is only $F$-term supersymmetry breaking,
their masses are subleading in an expansion in the breaking parameter $F/M^2$, because they arise from the operator
\beq
\C{L} \supset \int d^2\theta \; \frac{1}{M^3}  \Sigma^a W^{a,\alpha} \ov{D}^2 D_\alpha ( X^\dagger X) 
\eeq 
where $\Sigma^a$ is the adjoint chiral superfield whose fermion pairs with the gauginos, $M$ is the messenger mass scale and $\bra X \ket = \theta^2 F$ can preserve $R$-symmetry. The mass is thus $\C{O}(F^2/M^3)$. Actually this is an \emph{improvement} on models of Majorana gauginos based on many ISS or O'raifeartaigh models, where, even when $R$-symmetry is broken, the gaugino mass is \emph{third} order in the expansion~\cite{Izawa:1997gs,Kitano:2006xg,Csaki:2006wi,Abel:2007jx,Abel:2007nr,Abel:2008gv} unless there is further metastability at tree-level~\cite{Komargodski:2009jf,Giveon:2009yu,Dudas:2010qg}. 
One solution for both the Majorana and Dirac cases is to have a low messenger scale where $F \lesssim M^2$ (e.g. \cite{Amigo:2008rc}), but most work on Dirac gauginos has used instead D-terms~\cite{Fox:2002bu,Chacko:2004mi,Matos:2009xv,Benakli:2010gi,Carpenter:2010as,Dumitrescu:2010ca}, where the masses arise from the supersoft operator $\C{L} \supset \int d^2\theta \; \frac{1}{M} \Sigma^a W^{a,\alpha} W^\prime_{\alpha}$
where $\bra W^\prime_\alpha \ket = \theta_\alpha D$ is a D-term spurion. 

An alternative to this would be instead to suppress the scalar masses, using the screening that naturally occurs in gaugino mediation; then the suppression of the leading term would be irrelevant. To this end we present in section \ref{sec:generalmodel} a toy model of deconstructed \emph{ Dirac} gaugino mediation. This provides a generic phenomenological framework for implementing $F$-term supersymmetry breaking with Dirac gaugino masses. In deconstructed gaugino mediation \cite{ArkaniHamed:2001ca,Csaki:2001em,Cheng:2001an,DeSimone:2008gm,McGarrie:2010qr,Auzzi:2010mb,Sudano:2010vt,Auzzi:2010xc,McGarrie:2011dc}, the visible gauge groups couple via link-fields to a hidden gauge group which in turn couples to messengers and so to the supersymmetry breaking sector. The link-fields develop a vacuum expectation value $\mu_\ell$ and higgs the visible and hidden groups to the MSSM gauge groups at a scale below the messenger scale $M$. This \emph{screens} the two-loop scalar masses by a factor of $\mu_\ell/M$, but the visible gauginos are a linear combination of the gauginos from the two original gauge groups, and their masses are not suppressed (thus imitating the spectrum of the original higher-dimensional gaugino mediation \cite{Kaplan:1999ac,Chacko:1999mi}). If the ratio $\mu_\ell/M$ is sufficiently small, then the two-loop MSSM sfermion masses, given by 
\beq
m_{\tilde{f}}^{2-loop} \approx \sum_f C_2 (f, r) \frac{\sqrt{2} g^2_r}{16\pi^2} \frac{|F|}{M} \frac{\mu_{\ell}}{M}\, ,
\eeq
can be smaller than the three-loop contribution that comes from integrating out the gauginos 
\beq
m_{\tilde{f}}^{3-loop} \approx \sum_f C_2 (f, r)  \frac{g_r m_D }{2\pi} \sqrt{\log [\frac{m_R^2}{m_D^2}]}
\eeq
where $m_D$ is the Dirac gaugino mass and $m_R$ is the mass of the real component of the adjoint scalar. The latter appears at the leading order, and so the logarithm can provide a significant enhancement. It is important to realise that these three-loop contributions remain unscreened and are always present.

In Section \ref{sec:completion} we will show that this link-field framework sits very comfortably in a UV completion based on the ISS model, that both includes supersymmetry breaking and provides the additional adjoint degrees of freedom for the Dirac gaugino. It is  related to the recent work of \cite{Green:2010ww} which used strong dynamics to provide a deconstructed Majorana gaugino mediation model. In that case, as mentioned above, even with broken $R$-symmetry the Majorana gaugino masses are third order in $F/M^2$, but this suppression could be overcome by the screening. We shall argue in Section \ref{sec:completion} that it is even more natural to consider Dirac gauginos in this context, because the Dirac masses are naturally heavier than their Majorana counterparts and there is no required breaking of $R$-symmetry. 

There are three further issues associated with Dirac gaugino masses that our construction allows us to address. The first is that of the adjoint scalar masses, which in the context of minimal gauge mediation turn out to be tachyonic. This is because there are two types of mass terms, given by
\begin{align}
\C{L} \supset& -m_{\Sigma}^2 \Sigma^a \ov{\Sigma}^a - \frac{B_{\Sigma}}{2} ( \Sigma^a \Sigma^a +\ov{\Sigma}^a \ov{\Sigma}^a )\nonumber \\
\supset& -\frac{1}{2} (m_{\Sigma}^2 + B_{\Sigma}) | \Sigma^a + \ov{\Sigma}^a|^2 - \frac{1}{2} (m_{\Sigma}^2 - B_{\Sigma}) | \Sigma^a - \ov{\Sigma}^a|^2
\end{align}
and typically we find $B_{\Sigma} > m_{\Sigma}^2$. This problem is solved in the toy model of Section \ref{sec:generalmodel} by a judicious choice of adjoint couplings to the messengers. In the UV completion of section \ref{sec:completion} the couplings are more constrained, but there exists a different and rather natural solution: K\"ahler potential terms that are generically induced by the strong dynamics. These are able to lift the erstwhile tachyonic directions. The second problem is unification, which can be solved by splitting the messenger masses \cite{Benakli:2010gi} or adding ``bachelor'' states that complete the adjoint fields into a broken GUT adjoint multiplet \cite{Fox:2002bu}. In our setup, this can acquire a new solution due to the higgsing of two groups: although we do not examine the issue in great detail we argue that a form of dual and/or deflected unification should be possible~\cite{Abel:2008tx,Abel:2009bj}. The final problem is that of scalar tadpoles; since the hypercharge adjoint field is a singlet, in principle it can acquire a dangerous tadpole term in the potential. However, this does not occur in either Section~\ref{sec:generalmodel}  since the couplings to the messengers respect $\SU{5}$ (and can be chosen to cancel the tadpole even if this is not true), or in Section~\ref{sec:completion} because the adjoint comes from an $\SU{5}$ which is unbroken at the messenger scale!

Finally, we also briefly comment on the fact that there must be $R$-symmetry breaking in a hidden sector in order to cancel the cosmological constant in supergravity, which generates a small Majorana gaugino mass term through anomaly mediation. This will result in pseudo-Dirac rather than pure Dirac gauginos. This may have experimental consequences; although we do not consider discuss them here, focussing instead on understanding the spectrum of masses, there is a growing body of work on possible signals for Dirac gauginos \cite{SIGNALS}.

\section{Two general remarks}
 
\subsection{The connection with extra dimensions}

\begin{figure}
\centering{}\includegraphics[scale=0.35]{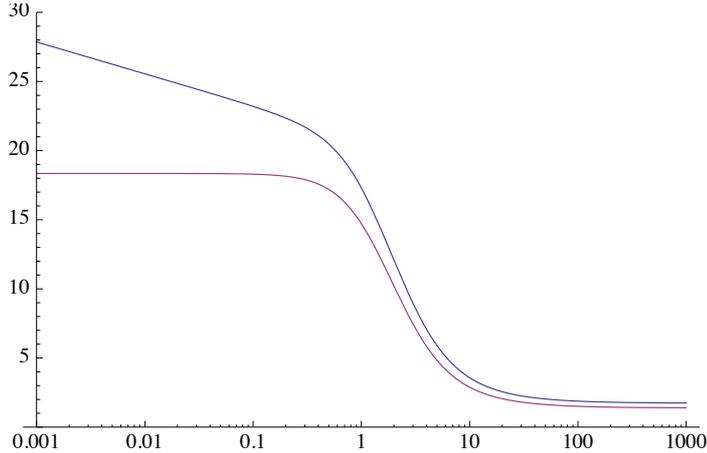}
\parbox{13cm}{
\caption{
\label{fig:gammas}
\emph{The ratio ${8\pi^{2}}{\Pi}/{M_{\lambda}^{2}}$
where the scalar mass-squareds are $m_{i}^{2}=4g^{2}C(R_{i})\Pi$,
in a slice of AdS$_5$ with supersymmetry breaking on the IR brane (taken from Ref.\cite{Abel:2010vb}).
The upper line is the zero momentum value, the lower line is the value
at the messenger scale. The lightest gaugino mode varies continuously
from Majorana to pure Dirac as the relative supersymmetry breaking
on the IR brane, $\xi$, increases. Note from the figure that the Majorana mass term is logarithmically 
renormalized (hence the difference between the two lines) while the (finite) Dirac mass term
is not.} } }
\end{figure}

Although this paper is not in the main concerned with extra-dimensions,
it is worth noting an extra-dimensional indication that
this kind of simple $R$-symmetric configuration is possible.
The model of Ref.\cite{Green:2010ww} supposedly corresponds to 
the deconstruction of a 5D gaugino mediation model (with only three nodes) 
in which supersymmetry breaking is mediated by bulk gauge modes. 
However in the simplest 5D models of gaugino mediation (in
which supersymmetry is broken by twisted -- Neumann/Dirichlet -- boundary
conditions) the lowest lying Weyl gaugino pairs up with a Kaluza-Klein
mode to form a massive pure \emph{Dirac} state, not a Majorana one
\cite{Gherghetta:2000kr, Marti:2001iw,Abel:2010vb}. \emph{Pseudo-Dirac} masses can result if supersymmetry
is instead broken by a non-zero gaugino mass-term located on one of the branes
\cite{Marti:2001iw,Abel:2010vb}. This was analyzed recently in Ref.\cite{Abel:2010vb} for a slice
of AdS$_{5}$ (which allows for an appealing interpretation in terms
of strongly coupled 4D field theories via the usual gauge/gravity
correspondence). The resulting 4D gaugino masses are reproduced for
reference in Figure \ref{fig:gammas}. The parameter $\xi$ represents
the relative size of the $F$-term VEV on the IR brane (the precise
definition, contained in \cite{Abel:2010vb}, doesn't matter here). At small
values of SUSY breaking the lowest lying gaugino state is Majorana,
but as $\xi$ is increased this state mixes with the lowest lying
KK mode to become pseudo-Dirac. In the $\xi\rightarrow\infty$ limit
the bulk gaugino wave-function is completely repelled from the IR
brane and effectively has a Dirichlet boundary condition there: the
resulting pure Dirac mass coincides with the twisted boundary condition
value in this limit. In the light
of these more general possibilities in extra dimensional models, it
indeed seems plausible that simple dynamically realised 4D configurations
should also be able to accommodate
both Dirac and pseudo-Dirac gauginos, along with the associated 
$R$-symmetry. 

\subsection{$R$-emergency}\vspace{-0.85cm}\hspace{4.4cm}{\footnote{The correct -- and less whimsical -- expression would be ``$R$-emergentness''.}}\\

\label{sec:R-emergency} 
\noindent  It is also worth elaborating a little on why $R$-symmetry
is naturally an \emph{emergent} phenomenon. Let us consider the example of Ref.\cite{MN}; 
the model was based on SQCD in the free-magnetic window (i.e. the ISS model). Ref.\cite{MN}
suggested that massive messenger fields $f$, $\tilde{f}$ would couple
to the ISS model at leading order as \begin{equation}
W^{(el)}=m_{Q}Q\tilde{Q}+\frac{1}{M_{X}}f\tilde{f}Q\tilde{Q}+M_{f}f\tilde{f}\end{equation}
where $Q$, $\tilde{Q}$ are the quarks of the electric SQCD theory
and $M_{X}$ is the scale of fundamental physics (the string or Planck
scale, say) at which the operator is generated. ISS makes use of Seiberg
duality to derive a magnetic low energy description in which the electric
quarks are confined into bound-state mesons $\Lambda\varphi\approx Q\tilde{Q}$,
where $\Lambda$ is the dynamical scale of the electric theory. The superpotential of the magnetic theory then has an operator
\begin{equation}
W^{(mag)}\supset\frac{\Lambda}{M_{X}}f\tilde{f}\varphi\end{equation}
in which $\Lambda/M_{X}$ appears as a suppressed Yukawa coupling.
As advertised the magnetic theory has an {}``almost $R$-symmetry''.
Indeed the classical superpotential of the magnetic theory is of the
form \begin{equation}
W^{(mag)}=hq\varphi\tilde{q}-\mu_{ISS}^{2}\varphi+M_{f}f\tilde{f}+\frac{\Lambda}{M_{X}}f\tilde{f}\varphi,\label{eq:wmag1}\end{equation}
where $\mu_{ISS}^{2}=\Lambda m_{Q}$. The first piece is essential
in Seiberg duality to make the moduli spaces of the electric and magnetic
theories match. Without the $\mu_{ISS}$ coupling and the small coupling
$\Lambda/M_{X}$ the theory has an anomaly-free $R$-symmetry under
which $R_{f}=R_{\tilde{f}}=1$. However the $\mu_{ISS}$ coupling
leaves only an anomalous $R$-symmetry (with $R(\varphi)=2$ and $R(q\tilde{q})=0$).
A non-perturbative term (which we do not show) is therefore induced
that breaks the $R$-symmetry explicitly. This leads to the phenomenon
(observed by ISS) that supersymmetry is broken in a metastable vacuum,
and there exist global supersymmetric minima, due to the anomalous
nature of the remaining $R$-symmetry. The fourth piece breaks all
$R$-symmetry explicitly, however the $\Lambda/M_{X}$ coupling can
be small enough to prevent the decay of the metastable minimum within
the lifetime of the universe. Indeed in accord with the Nelson-Seiberg
theorem a global superymmetric minimum appears in the messenger direction,
but it is parametrically far away in field space, namely where $\left\langle f\tilde{f}\right\rangle =M_{X}\mu_{ISS}^{2}/\Lambda$
and $\left\langle \varphi\right\rangle =-M_{f}M_{X}/\Lambda$.

Now, an important aspect of the above theory is that in the IR it
flows to a (trivial) fixed point. There are several theorems governing
the flow to conformal fixed points, the most important being that
the superconformal algebra relates operator dimensions to the $R$-charges
under a conserved $R$-symmetry known as the \emph{exact} $R$-symmetry
(see \cite{Strassler:2005qs} for a review): the relation is \begin{equation}
\dim\mathcal{O}=\frac{3}{2}R_{\mathcal{O}}\end{equation}
so that if an operator has $R_{\mathcal{O}}>2$ it is irrelevant at
the fixed point, and vice versa. There is a second theorem relevant
for our discussion: the smallest dimension that a spin-zero gauge invariant
operator can have is unity. Whenever the dimension of an operator
hits unity it becomes a free-field and decouples%
\footnote{In the event that the system of possible $R$-symmetries is underconstrained
(by the superpotential and by the vanishing of $\beta$-functions)
the exact $R$-symmetry can be determined by a third theorem, the
$a$-maximization theorem \cite{Intriligator:2003jj,Barnes:2004jj}: the exact $R$-symmetry is the
combination that maximizes $a(R)=3tr(R^{3})-tr(R)$ where the trace
is over fermions. %
}. 

These rules imply that there is a very general class of models, in
which massive weakly coupled messengers interact with a strongly coupled
supersymmetry breaking sector with an IR fixed point, that behave
exactly as the model of Ref.\cite{MN}. In particular, they allow
an emergent $R$-symmetry. (No reference to a magnetic dual is necessary.)
Suppose that all we know about the supersymmetry breaking sector is 
that it runs to an IR fixed point (free or interacting).  
The theory could be much more complicated than ISS, for example containing adjoints, 
product gauge groups, chiral superfields and so forth.
By gauge invariance, weakly coupled messenger superfields
$f$ and $\tilde{f}$ have to appear at leading order in the superpotential
as \begin{equation}
W^{(el)}=W_{SUSY-BREAKING}+\left(\sum_{i}\mathcal{O}_{i}\right)f\tilde{f}+M_{f}f\tilde{f},\end{equation}
where $\mathcal{O}_{i}$ are arbitrary additional operators involving
the supersymmetry breaking sector fields, and where we can safely
neglect terms higher order in $f\tilde{f}$ in what follows. Since
the minimum possible dimension of the $\mathcal{O}_{i}$ is unity,
their $R$-charges will be greater than $\frac{2}{3}$ (the free-field
value) and so there can be no exact $R$-symmetry compatible with
both the operators $\mathcal{O}_{i}f\tilde{f}$ and $M_{f}f\tilde{f}$.
Trivially though, one can always assign an $R$-charge to $f\tilde{f}$
that respects the exact $R$-symmetry if only $M_{f}f\tilde{f}$ is
present. Since the messenger fields are weakly coupled, their dimension
is by definition close to unity and $M_{f}f\tilde{f}$ is relevant,
whereas all the other operators $\left(\sum_{i}\mathcal{O}_{i}\right)f\tilde{f}$
are irrelevant becoming exactly marginal only when (as in Ref.\cite{MN}
at the trivial fixed point) the operator $\mathcal{O}_{i}$ hits the
unitarity bound and decouples as a free-field. The flow of the theory
is then inevitably towards restoring the exact $R$-symmetry compatible with the operator
$M_{f}f\tilde{f}$, which appears in the IR as an emergent symmetry.
Finally an operator such as $\mu_{ISS}^{2}$ can arrest the flow at
some low scale so that the erstwhile irrelevant operators are present
but suppressed%
\footnote{As we said, in the specific case of ISS the $\mu_{ISS}^{2}$ operator
explicitly breaks the exact $R$-symmetry but leaves intact an anomalous
$R$-symmetry which is a combination of the exact $R$-symmetry and
an anomalous $\U1_{A}$, leading to metastability.%
}. 

The simple model of Ref.\cite{MN} clearly conforms to this general
rule, however there are virtually limitless alternative possibilities.
For example it is possible for the strongly coupled sector to flow
close to an intermediate interacting fixed point, before becoming
overwhelmed by a relevant operator in $W_{SUSY-BREAKING}$ and flowing
to a new fixed point. At this stage one of the $\mathcal{O}_{i}$
may hit the unitarity bound and become a decoupled free-field $\mathcal{O}_{i}\equiv\varphi_{i}$.
If this happens repeatedly, as in a duality cascade for example, a number of suppressed $f\tilde{f}\varphi_{i}$
couplings could be generated.

\section{Deconstructed Dirac gaugino mediation}
\label{sec:generalmodel}

\subsection{Setup}

We now turn to an explicit realisation of Dirac gauginos in a simple model. This configuration can be considered as 
``toy'' in the sense that it includes neither a UV completion, nor dynamical supersymmetry breaking. In the following section we shall provide these two crucial ingredients by considering Seiberg duality, but for the moment let us concentrate on the phenomenological aspects.

\begin{figure}
\begin{center}
\epsfig{file=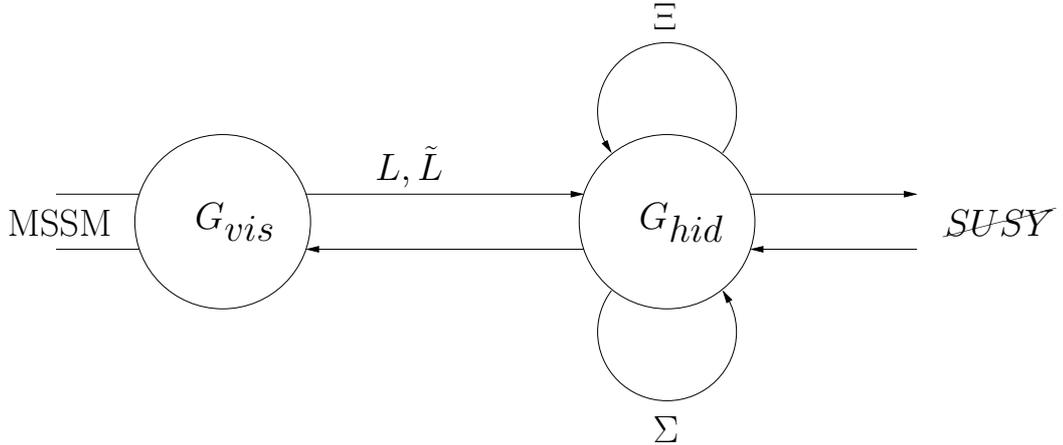,width=0.8\linewidth}
\caption{\em Deconstructed Dirac gaugino mediation in a two-site model\label{fig:setup}
}
\end{center}\end{figure}

The general setup is shown in Figure \ref{fig:setup}. It involves link-fields $L, \tilde{L}$ obtaining a VEV at a Higgsing scale $\mu_\ell$ via a supersymmetric mechanism, which breaks the $G_{vis} \times G_{hid}$ to the diagonal combination. The messengers of supersymmetry breaking are charged under the $G_{hid}$ group, as are two $G_{hid}$ adjoints, $\Sigma$ (which gives the gauginos Dirac masses), and an additional adjoint $\Xi$ which gives masses to certain link-fields.   This ``very deconstructed'' configuration is expected to provide the same gaugino mediation screening effect for the scalars noted in  Ref.\cite{Cheng:2001an}. Note that unlike the original version of deconstructed gaugino mediation \cite{Cheng:2001an} we require $G_{vis}$ to be identical to $G_{hid}$, otherwise after the Higgsing there remain additional massless ``bachelor'' states. We shall return to this point in detail in Section 4. Thus for $G_{vis} = \SU{3} \times \SU{2} \times \U1$ both indices of the link-fields fall into diagonal gauge blocks. 
The superpotential is 
\begin{align}
\label{canonical-dirac}
W =& W_{\rm MSSM} + W_{\rm higgsing} + W_{\rm mess} + W_{\cancel{SUSY}}
\end{align}
where 
\begin{align}
W_{\rm higgsing} =& K (\frac{1}{5} L \tilde{L} - \mu_\ell^2) + L \Xi\tilde{L} + m \Xi \Sigma
\end{align}
is essentially the higgsing superpotential of \cite{Cheng:2001an}, but with an optional additional mass $m$ coupling the two adjoints. We have suppressed gauge indices, but the term $m\Xi \Sigma$ should be understood as $2 m \tr \Xi\Sigma = m \Xi^{a} \Sigma^a$. $K$ is a Lagrange multiplier singlet field. We can choose whatever messenger sector we desire, however to generate Dirac gaugino masses we must also couple the messengers to the adjoint $\Xi$. This will then generate masses for the adjoint scalars. These could in principle be tachyonic, but by a judicious choice of couplings we can avoid such a disaster; as a concrete example with two pairs of messengers we choose:
\begin{align}
W_{\rm mess}=& S f_1 \tilde{f}_2 + M (f_1 \tilde{f}_1 +  f_2 \tilde{f}_2) + h_1 f_1 \Sigma \tilde{f}_1 + h_2 f_2 \Sigma \tilde{f}_2.
\end{align}
The $f_i, \tilde{f}_i$ are fundamental/anti-fundamental pairs  under $G_{hid}$ acting as  messengers, and $S$ is an $F$-term spurion. This falls into the class of models studied in \cite{Benakli:2008pg}; for $h_2 = -h_1$ the messenger couplings are essentially those of \cite{Amigo:2008rc}.  Notice that the superpotential preserves an $R$-symmetry ($R_K=R_S=R_\Xi=R_{f_2}=R_{\tilde{f}_1}=2$ and $R_\Sigma=R_{L}=R_{\tilde{L}}=R_{f_1}=R_{\tilde{f}_2}=0$) so that if gaugino masses are generated they will have to be Dirac. 

The $F$-term equation for $K$ causes the link-fields $L, \tilde{L}$ to acquire a VEV, breaking the gauge group to the diagonal combination. It is convenient to choose the VEVs to be equal so that $\bra L \ket = \bra \tilde{L} \ket = \mu_\ell$, but there is a global symmetry associated with the relative sizes of these VEVs that leads to a harmless goldstone boson (which could, however, be eaten by gauging or broken explicitly by other terms).  

In the more conventional picture of deconstructed gaugino mediation \cite{Cheng:2001an}, 
the gaugino of the second group is made to acquire a Majorana mass (with the help of additional $R$-violating operators), and upon higgsing the groups the lightest, diagonal, gaugino state acquires this same Majorana mass. However,  compared to \cite{Cheng:2001an}, our toy model has an additional adjoint field $\Xi$.
This field generates instead a Dirac mass for the second gaugino, and upon diagonalisation the lightest state is a pure Dirac gaugino. Indeed, the direct couplings between the messengers and the $G_{hid}$ group generate Dirac gaugino and adjoint scalar masses at the messenger scale of 
\begin{align}
\label{mdirac}
m_{D} =& I_f \frac{(h_1 - h_2) g_2}{\sqrt{2}} \frac{1}{16\pi^2}\frac{|F|^2}{6 M^3}  \nn\\
m_{\Sigma}^2 =& I_f \frac{1} {16\pi^2} \frac{|F|^2}{6 M^2} |h_1 - h_2|^2 \nn\\
B_{\Sigma} =& - I_f \frac{1} {16\pi^2} \frac{|F|^2}{3 M^2} (h_1^2 + h_2^2 + h_1 h_2)
\end{align}
where $I_f$ is the Dynkin index of the messengers under the appropriate group. Note that the indices on the $h_{i=1,2}$ refers to the messenger field, whereas the indices on the $g_{A=1,2}$ refer to the gauge node. As we said, depending on the relative size of $h_1$ and $h_2$ the supersymmetry breaking masses for the adjoint scalars could be tachyonic. However, it is clear that if we take $h_1$ and $h_2$ to be real with opposite signs for example then this will not be the case provided $h_2 < (\sqrt{3} - 2) h_1$. 

\label{unif-discussion}

There are interesting alternative schemes for gauge coupling unification in this model. In the context of gauge mediation models of Dirac gauginos, this is typically a problem because adding adjoints of $\SU{3}$ and $\SU{2}$ spoils unification, and extra fields must usually be added to restore it. Other than abandoning $\SU{5}$ unification, one approach is to add the remaining ``bachelor'' fields so that the adjoints sit in a $\mathbf{24}$ \cite{Fox:2002bu}. However, with messengers there is a tension with perturbativity at the GUT scale. The alternative is that the messengers themselves restore unification, with viable explicit models discused in \cite{Benakli:2010gi}. However, here there are very different possibilities. For example, below the higgsing scale, $G_{vis}$ is coupled to the MSSM fields plus chiral adjoint fields, but above this scale it couples only to the link-fields. In the toy model this leads to the visible $\SU{3}$ walking (with zero one loop beta function) up to the GUT scale. Furthermore, the three gauge groups obtain shifts introduced at the higgsing scale: the tree-level matching condition in the generic link-field configuration is
\beq
\frac{1}{\alpha^i_{diag}} = \frac{1}{\alpha^i_{vis}} + \frac{1}{\alpha^i_{hid}}.
\eeq
The general problem is then to preserve apparent unification in the diagonal coupling. There is always the possibility that both ${\alpha^i_{hid}}$ and ${\alpha^i_{vis}}$ unify, but for example if 
$\alpha^i_{vis}$ is very large, then diagonal gauge couplings are dominated by ${\alpha^i_{hid}}$. We then require unification in only the hidden gauge group in this limit! More generally, we see that there is the possibility of both dual and deflected unification, or indeed a combination of the two~\cite{Abel:2008tx,Abel:2009bj,singlets}. We leave this as an interesting topic for further study.

\subsection{Higgsing}

Let us now determine the lightest Dirac mass in detail beginning with the Higgsing. We define the mixing angle $\tan \vartheta = g_2/g_1$ and the linear combinations 
\begin{equation}
A_\pm  = \cos\vartheta A_{2,1} \pm \sin\vartheta A_{1,2} \, .
\end{equation}
Note that, assuming $\bra L \ket = \bra \tilde{L} \ket = \mu_\ell$,  the mass of the broken group $A_-$ is given by the term
\begin{align}
\C{L} \supset& |D_\mu L|^2 + |D_\mu \tilde{L}|^2  \nn\\
\supset & 2 (g_1^2 + g_2^2) \mu_\ell^2 \frac{1}{2} A_-^a A_-^a\, ,
\end{align} 
so we generate an $A_-$ gauge boson mass of $M_A= \sqrt{2 (g_1^2 + g_2^2)}\mu_\ell$. Fundamental fields of group 1 will be our standard model fields. They contain a term $g_1 A_1$ in their covariant derivatives, and thus couple to $A_+$ with strength $\frac{g_1g_2}{\sqrt{g_1^2 + g_2^2}}=\sin \vartheta g_1 $. This becomes the new coupling. Note that (as we noted above) if $g_1 \gg g_2$ the coupling strength of the diagonal group will be approximately $g_2$.

The covariant derivative of $L$, defining the first index to be the fundamental of group $1$, and the second as the antifundamental of group $2$, is given by
\begin{align}
D_\mu L_{i\tilde{j}} =& \partial_\mu L_{i\tilde{j}} + ig_1 A^a_{1, \mu} T^a_{ik} L_{kj} - ig_2 A^b_{2, \mu} T^b_{\tilde{k}\tilde{j}} L_{i\tilde{k}} \nn\\
=& \partial_\mu L_{i\tilde{j}} + i g_1 \sin\vartheta  [A_{+\mu},  L] +  
( i g_1 \cos\vartheta A_{-\mu} L  + i g_2\sin\vartheta  L A_{-\mu})  \nn\\
\rightarrow& \partial_\mu L_{i\tilde{j}} + \frac{i g}{\sqrt{2}} [A_{+\mu}, L ] + \frac{ig}{\sqrt{2}} \{A_{-\mu}, L \},
\end{align}
where in the last line we take $g_1 = g_2 \equiv g$.
Similarly, defining the first index of $\tilde{L}$ to be the fundamental of group $2$ and the second as the antifundamental of group $1$ we find (for $g_1 = g_2 \equiv g$)
\begin{equation}
D_\mu \tilde{L}_{i\tilde{j}} 
= \partial_\mu \tilde{L}_{i\tilde{j}} + \frac{ig}{\sqrt{2}} [A_{+\mu}, \tilde{L}] - \frac{ig}{\sqrt{2}} \{A_{-\mu}, \tilde{L} \}\, .
\end{equation}
Then note that if we take $L_{\pm} \equiv \frac{1}{\sqrt{2}} (L \pm \tilde{L})$ we have
\begin{align}
D_\mu L_\pm =& \partial_\mu L_\pm + ig_1 \sin\vartheta  [A_{+\mu}, L_\pm ] 
\nn \\
& \,\,\,\,\, +  \frac{(g_1\cos\vartheta  - g_2\sin\vartheta)}{2} [ A_-,L_\pm]  + \frac{(g_1\cos\vartheta  + g_2\sin\vartheta)}{2}  
\{ A_-, L_\mp \}\, .
\end{align}
The significance of this is that only $L_+$ obtains a VEV. Moreover, it is $L_+$ that obtains a mass from the term $W \supset \frac{1}{5} K L \tilde{L} + L A \tilde{L}$ in the superpotential; the scalar $L_-$ degrees of freedom are either eaten by the broken gauge group or given a mass by the D-terms (except the trace component corresponding to the goldstone boson of the spontaneously broken global $U(1)$). 

We turn now to the fermion masses, denoting the fermionic components of the superfields as $L, \tilde{L}\rightarrow \eta, \tilde{\eta}$; $K\rightarrow \chi$; $\Sigma \rightarrow \varsigma$; $\Xi \rightarrow \xi$. Writing $L = (\mu_\ell + l) \delta_{ij}  + \sqrt{2} \ell^a T^a, \tilde{L} = (\mu_\ell + \tilde{l}) \delta_{ij} + \sqrt{2} \tilde{\ell}^a T^a$, (where the $\sqrt{2}$ factors are so that the kinetic terms are correctly normalised), the fermion masses arise through \begin{align}
\C{L} \supset& - \sqrt{2} \mu_\ell \chi \eta_+ - m \xi \varsigma - 2\mu_\ell  \tr (\xi \eta_+) \nn\\
&-  g_1 \sqrt{2} \tr (\lambda_1 \eta L^*) +  g_1 \sqrt{2} \tr ( \tilde{\eta} \lambda_1  \tilde{L}^*) + g_2 \sqrt{2} \tr (\eta \lambda_2  L^*) - g_2 \sqrt{2} \tr (  \lambda_2 \tilde{\eta} \tilde{L}^*) \nn\\
= & -\sqrt{2} \mu_\ell (\chi \eta_+) - m (\xi^a \varsigma^a ) - \mu_\ell (\xi^a \eta_+^a) - M_A (\lambda_-^a \eta_-^a)\, .
\end{align}
Note that these are the same masses as the scalar counterparts, since the masses are generated in a supersymmetric way. (There remains a massless fermionic superpartner of the massless scalar singlet, $l_-$, corresponding to the goldstone boson of the global $\U1$ rotating the $L, \tilde{L}$.) 
To the above we can now add the supersymmetry breaking Dirac mass term $-m_D \varsigma \lambda_2 = -m_D \cos\vartheta (\varsigma \lambda_+)  + m_D \sin\vartheta (\varsigma \lambda_-)  $ generated by the coupling of the adjoint to the messengers.  The entire mass matrix for the vector of adjoint fermions 
$(\xi,\varsigma,\eta_+,\eta_-,\lambda_+,\lambda_-)^a$ takes the form 
\beq
m_{\psi_{Adj}} = \frac{1}{2}\left( 
\begin{array}{cccccc}
0 & m & \mu_\ell & 0 & 0 & 0 \nn \\
m & 0 & 0 & -m_D \sin\vartheta & m_D \cos\vartheta & 0 \nn \\
 \mu_\ell & 0 &0 & 0 & 0 & 0 \nn \\
0 & -m_D \sin\vartheta & 0 & 0 & 0 & M_A \nn \\
0 & m_D \cos\vartheta & 0 & 0 & 0 & 0 \nn \\
0 & 0 & 0 & M_A & 0 & 0 
\end{array}
\right)
\eeq
This leaves the lightest state as the Dirac gaugino being composed of $\lambda_+$ and the linear combination $\mu_\ell\, \varsigma -m  \, \eta_+ $, with mass given by 
\beq
m_\lambda = m_D \frac{\mu_\ell}{\sqrt{2(m^2 + \mu_\ell^2)}} - 
\frac{ m_D^3 \mu_\ell}{(m^2 + \mu_\ell^2)^{5/2}} \left(\frac{m^2 \mu_\ell^2 +M_A^2 m^2 + \mu_\ell^4}
{4\sqrt{2}M_A^2}\right) +  \C{O}\bigg(\frac{m_D^5}{\mu_\ell^4}\bigg).
\eeq
Note that the supersymmetric masses for the adjoints arise from the terms
\begin{align}
W \supset& L\Xi\tilde{L} + m \Xi \Sigma \nn\\
\rightarrow -\C{L} \supset& | \mu_\ell L_+^a + m \Sigma^a|^2 + (m^2 + \mu_\ell^2)|\Xi^a|^2 
\end{align}
and thus there is a combination of $L_+$ and $\Sigma$ that is massless at the supersymmetric level. Once supersymmetry is broken, we find that the lightest adjoint scalar states have mass squared approximately $\frac{\mu_\ell^2}{m^2 + \mu_\ell^2} (m_\Sigma^2 \pm B_\Sigma) $ where $m_{\Sigma}, B_\Sigma$ are the one-loop masses generated above in equation (\ref{mdirac}). 

\subsection{Scales}

We can then compare the soft masses in the visible sector. Let us for simplicity take $h_1 = - h_2 \equiv h$. Then the Dirac gaugino masses are
\beq
m_{\lambda} \simeq I_f \sqrt{2} h g_r \frac{1}{16\pi^2}\frac{|F|^2}{6 M^3}  \frac{\mu_\ell}{\sqrt{2(m^2 + \mu_\ell^2)}}\, ,
\eeq
the two-loop sfermion masses are
\beq
m_{\tilde{f}}^{2-loop} \approx \sum_f C_2 (f, r) \frac{\sqrt{2} g^2_r}{16\pi^2} \frac{|F|}{M} \frac{\mu_{\ell}}{M}\, ,
\eeq
and the three-loop sfermion masses are 
\beq
m_{\tilde{f}}^{3-loop} \approx \sum_f C_2 (f, r) I_f \sqrt{2} h g_r^2 \frac{\mu_\ell}{\sqrt{2(m^2 + \mu_\ell^2)}} \frac{1}{32\pi^3}\frac{|F|^2}{6 M^3}\sqrt{\log [\frac{16\pi^2 M^2}{g^2 F} ]}\, .
\eeq
 Clearly the scalar masses are screened in the expected manner.
An extremely interesting feature of this configuration is that when the adjoint mass $m$ is less than $\mu_\ell$ it does not substantially disrupt the Dirac gaugino mass; we can vary it from zero to the order of the Higgsing scale without suppression. 
Conversely, when $M\sim m \gg \mu_\ell$ the screened masses are both of order $\frac{1}{16\pi^2 }\frac{F}{M^2} \mu_\ell$; only the ratio $\frac{F}{M^2}$ enters, and the absolute scale of supersymmetry breaking can be decoupled in this limit.

\section{A fully dynamical UV completion}
\label{sec:completion}

\subsection{Framework}
\label{subsec:model}

In this section we present a completely perturbative and well-controlled
UV completion. It is a strongly coupled SQCD theory that yields the toy model 
outlined in the previous section as its magnetic Seiberg dual, complete with ISS supersymmetry breaking. 
The model is closely related to that of 
Ref.\cite{Green:2010ww}, with the addition of three important (but natural) elements. First of all we add an elementary adjoint meson 
in the electric theory with the usual Yukawa coupling to the electric quarks which is however suppressed. This leads to the extra adjoint degree of freedom required for the Dirac gaugino in the magnetic theory.  
Second we include higher order operators in the K\"ahler potential of the magnetic theory, which are of a size consistent 
with their induction by the strongly coupling SQCD: these will be sufficient to lift and stabilise would-be tachyonic directions.
Finally we explicitly break some of the global flavour symmetries in the couplings: this is required to generate a non-zero Dirac mass (it is the equivalent of not having $h_1=h_2$ in the toy model).  
As we shall see these minor and perfectly consistent modifications yields a low energy model 
that mimics that of the previous section, with a massive 
Dirac gaugino as the lightest state. In addition 
 a controlled  breaking of $R$-symmetry generated as in \cite{MN} can make them arbitrarily pseudo-Dirac.

To describe the model in detail let us first return
to Seiberg duality of ordinary SQCD \cite{S:Duality,Intriligator:1995au}
and concentrate on the role played by mesons.
The familiar formulation of Seiberg duality is based on $\sunc$ with $\nf$ flavours of quarks and anti-quarks
and an empty electric superpotential, $W^{(el)}=0$. The flavour symmetry is \[
\sunf\times\sunf\times\U1_{B}\times\U1_{R},\]
with the particle content as shown in Table \ref{sqcd0}. We will
always be working in the window $\nc+1 <\nf<\frac{3}{2}\nc$. Under this assumption the gauge couplings 
diverge at some scale $\Lambda$ and one can dualize to an IR-free magnetic theory with the spectrum shown in Table
\ref{sqcd0-1}.

The two theories satisfy stringent tests of for example anomaly and baryon
matching, if one adds a superpotential to this theory of \begin{equation}
W^{(mag)}=h\, q\varphi\tilde{q}.\end{equation}
The parameter $h$ is difficult to compute (because it depends on
the non-holomorphic part of the theory, i.e. the Kahler potential). It is usually taken to be $\mathcal{O}\left(1\right)$, and to avoid clutter we shall set these couplings to 
$h=1$ until it is necessary to do otherwise. 
If one adds a quark mass term to the electric theory then one also recovers the
linear meson term of the magnetic superpotential shown in Eq.(\ref{eq:wmag1}). 

There is an equally valid reversed form of the duality, in which it is the
electric theory (shown in Table \ref{sqcd0-2}) that contains the
elementary mesons, $\Phi$, and has a non-empty superpotential,\begin{equation}
W^{(el)}=Q\Phi\tilde{Q}.\end{equation}

\noindent Upon dualizing, the quarks still get bound
into meson states in the magnetic theory, $\varphi\sim Q\tilde{Q}/\Lambda$. 
(Indeed the only way the 
superpotential would be able to change this behaviour would be if it contained quark mass terms larger than the dynamical scale $\Lambda$.) However the magnetic superpotential
is \begin{equation}
W^{(mag)}= q\varphi\tilde{q}+\Lambda\varphi\Phi.\end{equation}
Both sets of mesons $\varphi$ and $\Phi$ have mass $\sim\Lambda$
and can be integrated out of the theory below this scale, allowing us to identify $\Lambda\Phi\sim q\tilde{q}$,
and leading to a magnetic theory at low energy that has an empty superpotential
and the meson-free spectrum of Table \ref{sqcd0-1-1}. 

%\clearpage

\begin{table}[hpt]
\centering{}\begin{tabular}{|c||c|c|c|c|c|}
\hline 
$ $ & $\sunc$ & $\sunf$ & $\sunf$ & $\U1_{B}$ & $\U1_{R}$\tabularnewline
\hline
\hline 
$Q$ & $\fund$ & $\fund$ & 1 & $\frac{1}{\nc}$ & $1-\frac{\nc}{\nf}$\tabularnewline
\hline 
$\tilde{Q}$ & $\afund$ & 1 & $\afund$ & $-\frac{1}{\nc}$ & $1-\frac{\nc}{\nf}$\tabularnewline
\hline
\end{tabular}
\parbox{13cm}{\caption{\emph{Spectrum and anomaly free charges of the electric theory in
the canonical formulation of Seiberg duality}.\label{sqcd0}}}
\end{table}

\begin{table}[hp]
\centering{}\begin{tabular}{|c||c|c|c|c|c|}
\hline 
$ $ & $\sunnc$ & $\sunf$ & $\sunf$ & $\U1_{B}$ & $\U1_{R}$\tabularnewline
\hline
\hline 
$q$ & $\fund$ & $\afund$ & 1 & $\frac{1}{\nnc}$ & $1-\frac{\nnc}{\nf}$\tabularnewline
\hline 
$\tilde{q}$ & $\afund$ & 1 & $\fund$ & $-\frac{1}{\nnc}$ & $1-\frac{\nnc}{\nf}$\tabularnewline
\hline 
$\varphi$ & 1 & $\fund$ & $\afund$ & 0 & $2\frac{\nnc}{\nf}$\tabularnewline
\hline
\end{tabular}
\parbox{13cm}{\caption{\emph{Spectrum and anomaly free charges of the magnetic theory in
the canonical formulation of Seiberg duality.}\label{sqcd0-1}}}
\end{table}

\begin{table}[ph]
\centering{}\begin{tabular}{|c||c|c|c|c|c|}
\hline 
$ $ & $\sunc$ & $\sunf$ & $\sunf$ & $\U1_{B}$ & $\U1_{R}$\tabularnewline
\hline
\hline 
$Q$ & $\fund$ & $\fund$ & 1 & $\frac{1}{\nc}$ & $1-\frac{\nc}{\nf}$\tabularnewline
\hline 
$\tilde{Q}$ & $\afund$ & 1 & $\afund$ & $-\frac{1}{\nc}$ & $1-\frac{\nc}{\nf}$\tabularnewline
\hline 
$\Phi$ & 1 & $\afund$ & $\fund$ & 0 & $2\frac{\nc}{\nf}$\tabularnewline
\hline
\end{tabular}
\parbox{13cm}{\caption{\emph{Spectrum and anomaly free charges of the electric theory in
the reversed formulation of Seiberg duality}.\label{sqcd0-2}}}
\end{table}

\begin{table}[ph]
\centering{}\begin{tabular}{|c||c|c|c|c|c|}
\hline 
$ $ & $\sunnc$ & $\sunf$ & $\sunf$ & $\U1_{B}$ & $\U1_{R}$\tabularnewline
\hline
\hline 
$q$ & $\fund$ & $\afund$ & 1 & $\frac{1}{\nnc}$ & $1-\frac{\nnc}{\nf}$\tabularnewline
\hline 
$\tilde{q}$ & $\afund$ & 1 & $\fund$ & $-\frac{1}{\nnc}$ & $1-\frac{\nnc}{\nf}$\tabularnewline
\hline
\end{tabular}
\parbox{13cm}{\caption{\emph{Spectrum and anomaly free charges of the magnetic theory in
the reversed formulation of Seiberg duality.}\label{sqcd0-1-1}}}
\end{table}

\clearpage

\begin{figure}
\begin{center}
\epsfig{file=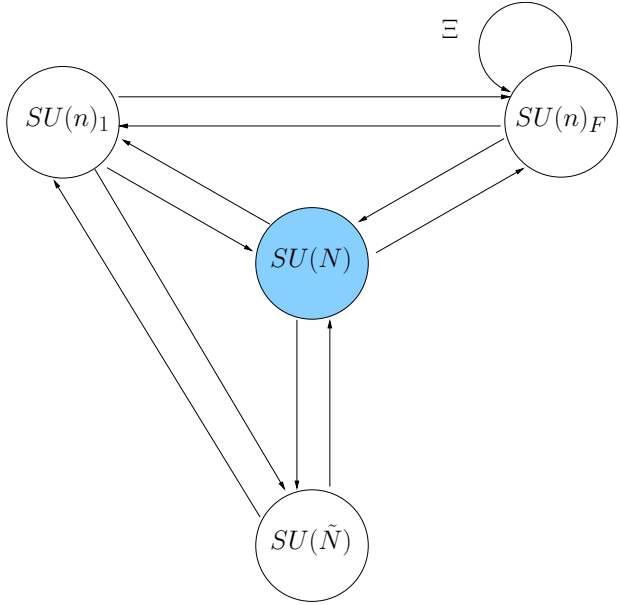,width=0.5\linewidth}
\parbox{13cm}{
\caption{\em \label{compuv}The UV completion of the model. The central node of the quiver is the dualizing ``colour'' group; The external nodes are flavours that we gauge to give the Standard Model.}
}
\end{center}
\end{figure}

\begin{figure}
\begin{center}
\epsfig{file=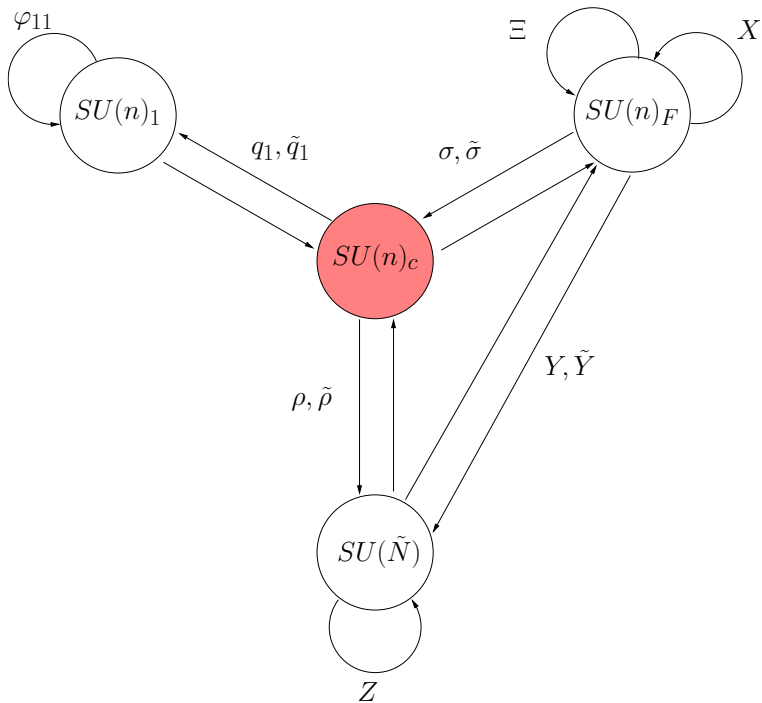,width=0.6\linewidth}
\parbox{13cm}{
\caption{\em \label{compir} The magnetic Seiberg dual of the quiver in Figure~\ref{compuv}. Supersymmetry is broken by the rank-condition with the $\SU{\tilde{N} \equiv N-n}$ node providing the $F$-term breaking: $F_Z\neq 0$. }
}
\end{center}
\end{figure}

Clearly one can more generally add an arbitrary number of mesons in the electric
theory, together with the corresponding Yukawa coupling. These and their
counterpart magnetic mesons will then be integrated out of the magnetic
theory. The remaining unpaired mesons will still appear in the magnetic
theory. The sum of the elementary electric mesons, and the massless composite magnetic mesons 
is clearly $N_f^{2}$.
We can use this freedom to add couplings to the electric theory in a block diagonal configuration that explicitly breaks the flavour symmetry 
(into $\nnc\times\nnc$ and $(\nf-\nnc)\times(\nf-\nnc)=\nc\times\nc$
blocks), so that the remaining magnetic mesons are of the form 
\begin{equation}
\varphi\Rrightarrow\left(\begin{array}{cc}
\varphi_{11} & -\\
- & \varphi_{22}\end{array}\right)\,\,\begin{array}{l}
\}\,\,\,\,\nnc\\
\}\,\,\,\,\nc\end{array}\end{equation}
while the electric ones are \begin{equation}
\Phi\Rrightarrow\left(\begin{array}{cc}
- & \Phi_{12}\\
\Phi_{21} & -\end{array}\right).\end{equation}
The flavour symmetries are broken to $\sunf\rightarrow\sunnc\times\sunc$
by this choice. The bar signifies that the mesons are
absent from the theory (not that they are zero). 
The quarks can also
be split into $\nnc\times\nnc$ and $\nc\times n$ blocks (where the
second index is colour); \begin{equation}
q\Rrightarrow\left(\begin{array}{c}
q_{1}\\
q_{2}\end{array}\right)\,\,\begin{array}{l}
\}\,\,\,\,\nnc\\
\}\,\,\,\,\nc\end{array}.\end{equation}
For the moment let us retain the maximum possible flavour symmetry
in the $\mu_{ISS}^{2}$ operators, so that the magnetic superpotential
takes the form \begin{equation}
W^{(mag)}= q_{1}\varphi_{11}\tilde{q}_{1}+ q_{2}\varphi_{22}\tilde{q}_{2}- \mu_{1}^{2}\varphi_{11}- \mu_{2}^{2}\varphi_{22}\end{equation}
where the $\mu_{1,2}^{2}$ terms again arise from flavour-diagonal
quark mass terms in the electric theory and where the Yukawa couplings are again all set to unity for the moment. 

Note that because the off-diagonal block is absent, the rank condition
of ISS factorizes. With the chosen configuration the rank condition
is saturated for the upper block which does not break supersymmetry,
while the lower block breaks
supersymmetry in the standard ISS manner. The states in the lower
block are therefore split again into entries corresponding to the
$n$ zero and $\nc-n$ non-zero $F_{\varphi}$-terms; \begin{equation}
\varphi_{22}\Rrightarrow\left(\begin{array}{cc}
X & Y\\
\tilde{Y} & Z\end{array}\right)\,\,\begin{array}{l}
\}\,\,\,\,\nnc\\
\}\,(\nc-\nnc)\end{array}\end{equation}
\begin{equation}
q_{2}\Rrightarrow\left(\begin{array}{c}
\sigma\\
\rho\end{array}\right)\,\,\begin{array}{l}
\}\,\,\,\,\nnc\\
\}\,(\nc-\nnc)\end{array}.\end{equation}
Every $F$-term that vanishes corresponds to non-zero quark VEVs;\begin{eqnarray}
\left\langle q_{1}\tilde{q}_{1}\right\rangle  & = & \mu_{1}^{2}\nonumber \\
\left\langle \sigma\tilde{\sigma}\right\rangle  & = & \mu_{2}^{2}\nonumber \\
\left\langle \rho\tilde{\rho}\right\rangle  & = & 0\, .\label{eq:mu}\end{eqnarray}
These VEVs break the $\sunnc_{F}\times\sunnc_{c}$ flavour/colour
symmetry of the $\sigma,\tilde{\sigma},X$ block to its diagonal subgroup,
which we refer to as $\sunnc_{\sigma}$. This group is orthogonal
to the flavour group of the upper block, which we refer to as $\sunnc_{1}$.
We therefore find a product of three nonabelian factors,
together with two baryon numbers and the exact $R$-symmetry; $\sunnc_{1}\times\sunnc_{\sigma}\times\SU{\nc-\nnc}_{\rho}\times\U1_{B}\times\U1_{B'}\times\U1_{R}$.
(The additional $\U1_{B'}$ factor is in the centre of the parent
$\sunf$ symmetry that we have broken by hand with our choice of meson
assignment.) 

We now further break the flavour symmetries: for our purposes the flavour symmetries have to be weakly
gauged for the first $\sunnc\times\sunnc$ factors so let us now move to a more general theory consistent with 
this, in order to avoid massless Goldstone modes associated with the spontaneous breaking of global symmetry. 
We can do this by splitting the  $\mu_{2}^{2}$ operator, so that \begin{equation}
\mu_{ISS}^{2}\Rrightarrow\left(\begin{array}{ccc}
\mu_{1}^{2} & 0 & 0\\
0 & \mu_{2}^{2} & 0\\
0 & 0 & \mu_{3}^{2}\end{array}\right)\,\,\begin{array}{l}
\}\,\,\,\,\nnc\\
\}\,\,\,\,\nnc\\
\}\,(\nc-\nnc)\end{array}.\end{equation}
In order to keep the breaking pattern of Eq.(\ref{eq:mu}), or equivalently
to avoid tachyons, we require $\mu_{3}^{2}\leq\mu_{2}^{2}$ as we shall shortly see. Later we will also be breaking the flavour symmetry in the Yukawa couplings.

Thus far, apart from the flavour breaking in the $\mu_i$'s, the set-up is as described in Ref.\cite{Green:2010ww}. We now come to our first important modification:
in the electric theory we add a meson state $\Xi$ which couples in the superpotential as,
\begin{equation}
W^{el} \supset h_\xi Q \Xi \tilde{Q}\, .
\end{equation} 
Just as the other mesons $\Phi_{12}$ and $\Phi_{21}$ were identified with bilinears of magnetic quarks, this meson  
would ultimately be identified as $\Xi\sim \sigma\tilde{\sigma}/h_\xi \Lambda$.
Indeed in the magnetic theory the Yukawa coupling becomes a mass term 
for the new adjoint $\Xi$ and the composite meson $X$, whose value is $m=h_\xi \Lambda $, and integrating out the massive states enforces this identification in the usual manner. 
However suppose that the coupling is much smaller than unity, $h_\xi \ll 1$ (which is perfectly acceptable).
Then  the mass is much less than $\Lambda$ and could be  comparable to or less than the $\mu_i^2$. We are not then 
entitled to integrate the states out and have to retain both of them in the magnetic theory. The full
renormalizable superpotential (without integrating out any degrees of freedom
due to the effects of the $\mu_{1}^{2}$ and $\mu_{2}^{2}$ couplings either)
is then \begin{equation}
W^{(mag)}=q_{1}\varphi_{11}\tilde{q}_{1}+ \sigma X\tilde{\sigma} + m \Xi X +\rho Z\tilde{\rho}+ \sigma Y\tilde{\rho}+ \rho\tilde{Y}\tilde{\sigma}-\mu_{1}^{2}\varphi_{11}-\mu_{2}^{2}X-\mu_{3}^{2}Z\, .\label{eq:wfin}\end{equation}

As for the K\"ahler potential of the magnetic theory, we will mostly assume it to be canonical (i.e. we neglect terms multiplied by factors of $\mu^2_i/\Lambda^2 $) except for an important contribution given by 
\begin{equation}
\delta K \supset \alpha_{flavour}^2 \frac{|Z|^2 |\Xi|^2}{\Lambda^2}\, .
\end{equation}
One expects these higher order operators to be be induced through the gauged flavour interactions of $\Xi$ to the strongly coupled sector, beginning at the one-flavour-loop level. (There are of course also more direct couplings through the $h_\xi$ coupling of $\Xi$ to the electric quarks, but these will turn out to be negligible by comparison.) This is our second important modification: ultimately we have $F_Z=\mu_3^2$ so this term generates a mass-squared term for the scalar $\Xi$ field of order 
\begin{equation}
\label{mkahler}
m^2_\Xi \sim \alpha^2_{flavour} \frac{\mu_3^4 }{\Lambda^2}\, .
\end{equation}
This will be instrumental in lifting tachyonic directions. 
Other higher order operators could appear, but they do not significantly change the behaviour. 

The charges under the diagonal $\sunnc_{1}\times\sunnc_{\sigma}\times\SU{\nc-\nnc}_{\rho}\times\U1_{B'}\times\U1_{B}\times\U1_{R}$
symmetry are shown in Table~\ref{blocked}, together with the residual
anomalous $R$-symmetry $\U1_{R'}$ that remains after the linear
meson terms are added. Clearly the new adjoint $\Xi$ and the corresponding mass term mimic the superfield of the toy model. 
The electric and magnetic models are represented graphically as quiver diagrams in Figures~\ref{compuv}
and \ref{compir}. 
\begin{table}
\centering{}\begin{tabular}{|c||c|c|c|c|c|c|c|}
\hline 
$ $ & $\sunnc_{1}$ & $\sunnc_{\sigma}$ & $\SU{\nc-\nnc}_{\rho}$ & $\U1_{B}$ & $\U1_{B'}$ & $\U1_{R}$ & $\U1_{R'}$\tabularnewline
\hline
\hline 
$q_{1}$ & $\fund$ & $\afund$ & 1 & $\frac{1}{\nnc}$ & $\frac{1}{\nnc}$ & $1-\frac{\nnc}{\nf}$ & 0\tabularnewline
\hline 
$\tilde{q}_{1}$ & $\afund$ & $\fund$ & 1 & $-\frac{1}{\nnc}$ & $-\frac{1}{\nnc}$ & $1-\frac{\nnc}{\nf}$ & 0\tabularnewline
\hline 
$\sigma$ & 1 & Adj & 1 & $\frac{1}{\nnc}$ & $-\frac{1}{\nnc}$ & $1-\frac{\nnc}{\nf}$ & 0\tabularnewline
\hline 
$\tilde{\sigma}$ & 1 & Adj & 1 & $-\frac{1}{\nnc}$ & $\frac{1}{\nnc}$ & $1-\frac{\nnc}{\nf}$ & 0\tabularnewline
\hline 
$\rho$ & 1 & $\fund$ & $\afund$ & $\frac{1}{\nnc}$ & $-\frac{1}{\nnc}$ & $1-\frac{\nnc}{\nf}$ & 0\tabularnewline
\hline 
$\tilde{\rho}$ & 1 & $\afund$ & $\fund$ & $-\frac{1}{\nnc}$ & $\frac{1}{\nnc}$ & $1-\frac{\nnc}{\nf}$ & 0\tabularnewline
\hline
\hline 
$\varphi_{11}$ & Adj & 1 & 1 & 0 & 0 & $2\frac{\nnc}{\nf}$ & 2\tabularnewline
\hline 
$X$ & 1 & Adj & 1 & 0 & 0 & $2\frac{\nnc}{\nf}$ & 2\tabularnewline
\hline 
$Z$ & 1 & 1 & Adj & 0 & 0 & $2\frac{\nnc}{\nf}$ & 2\tabularnewline
\hline 
$Y$ & 1 & $\fund$ & $\afund$ & 0 & 0 & $2\frac{\nnc}{\nf}$ & 2\tabularnewline
\hline 
$\tilde{Y}$ & 1 & $\afund$ & $\fund$ & 0 & 0 & $2\frac{\nnc}{\nf}$ & 2\tabularnewline
\hline
\hline 
$\Xi$ & 1 & Adj & 1 & 0 & 0 & $2-2\frac{\nnc}{\nf}$ & 0\tabularnewline
\hline 
\end{tabular}
\parbox{13cm}{\caption{\emph{Spectrum and anomaly free charges of the magnetic theory in
the split meson model. The first
two non-abelian factors are weakly gauged. The exact $R$ symmetry
}$\U1_{R}$\emph{ is broken to the anomalous }$\U1_{R'}$\emph{ by
the linear meson terms. The superfield \, $\Xi$ is a ``residual electric meson'' which gets a mass term with $X$ that is much smaller than the dynamical scale: hence both fields remain in the spectrum.}\label{blocked}}}
\end{table}

It is easy to see that this model with the superpotential of Eq.(\ref{eq:wfin})
approximates the model of Section \ref{sec:generalmodel}.
Indeed let $\nnc=5$ so that we are working in the simplifying $\SU{5}$ GUT framework. 
Since the only non-vanishing $F$-term is $F_{Z},$ we can identify \begin{eqnarray}
Z & \equiv & S\nonumber \\
\rho & \equiv & f_{1}\nonumber \\
\tilde{\rho} & \equiv & \tilde{f}_{2}\nonumber \\
Y & \equiv & f_{2}\nonumber \\
\tilde{Y} & \equiv & \tilde{f}_{1}.\end{eqnarray}
The $\SU{n}_F\times\SU{n}_c $ group is higgsed and collapses into a single node with attached 
adjoints reproducing the quiver of the toy model. However note that with this identification the adjoint fields corresponding
to $\Sigma$ in the superpotential will be given by $\sigma$, $\tilde{\sigma}$ and $\Xi$,
not as one might have expected $X$, which has no equivalent in the toy model and which 
would in any case have the wrong $R$-charge. Finally the role of $K$ and the link-fields is played by $\varphi_{11}$
and $q_{1}$, $\tilde{q}_{1}$ respectively. One nice feature about
this model is that the mass terms $M(f_{1}\tilde{f}_{1}+f_{2}\tilde{f}_{2})$
are automatically generated by the VEVs of $\sigma$ and $\tilde{\sigma}$,
so that $M\sim\mu_{2}$. Note also that the adjoints themselves are then
excitations around these VEVs with mass $\mathcal{O}(g_{2}\mu_{2})$. 

In order to verify this last point, let us determine precisely how
$\sigma$, $\tilde{\sigma}$ and $\Xi$ end up playing the role of the $\Sigma$ adjoints -- i.e. first consider just the breaking that collapses the quiver diagram into the toy model. We shall denote the gauge couplings of the 
$\sunnc_{F}\times\sunnc_{c}$ factors $g_F$ and $g_c$, at the scale of symmetry breaking $\sim \mu_2$. 
The pattern of higgsing of these factors is very much as in Section 3.1. 
It is convenient to define the magnetic quark fields as 
\begin{equation}
q_{2}=\left(\begin{array}{c}
\mu_{2}+\frac{1}{\sqrt{2}}(\sigma_{+}+\sigma_{-})\\
\frac{1}{\sqrt{2}}(\rho_{+}+\rho_{-})\end{array}\right)\,\,\,;\,\,\tilde{q}_{2}=\left(\begin{array}{c}
\mu_{2}+\frac{1}{\sqrt{2}}(\sigma_{+}-\sigma_{-})\\
\frac{1}{\sqrt{2}}(\rho_{+}^{c}-\rho_{-}^{c})\end{array}\right)\end{equation}
where we have put a charge conjugation on the $\rho_{\pm}$ elements of $\tilde{q}_2$. 
Denoting the gauge fields for the $\sunnc_{F}\times\sunnc_{c}$
factors by $A_{F}^{a}$ and $A_{c}^{a}$ (where $a$ is the adjoint index),
and working with the linear combinations
\begin{equation}
A'_\pm  = \cos\vartheta' A_{c,F} \pm \sin\vartheta' A_{F,c} \, ,
\end{equation}
where $\tan\vartheta'=g_c/g_F$, 
one can check that, with the VEVs in Eq.(\ref{eq:mu}), the
combination $A'_-$ gets a mass while the combination $A'_+$
remains light and gauges the diagonal $\sunnc_{\sigma}$ as required. Thus we identify $A'_+\equiv A_2$ in the toy model 
and call the heavy combination $A'_-\equiv B_2$: the mass of $B_2$ is $M_B=\sqrt{2(g_F^2+g_c^2)}\mu_{2}$.

For the masses of the matter fields, we need the covariant derivatives of $\sigma$ which can be read off from the discussion of the toy model;  
\begin{align}
D_\mu \sigma_\pm =& \partial_\mu \sigma_\pm + i{g_F}\sin\vartheta'  [A_{2\, \mu}, \sigma_\pm ] \nn \\
& 
+ \frac{(ig_F\cos\vartheta' - ig_c\sin\vartheta' )}{{2}} [B_{2\, \mu}, \sigma_\pm ]
+ \frac{(ig_F\cos\vartheta' + ig_c\sin\vartheta' )}{{2}}  \{B_{2\,\mu}, \sigma_\mp \} \, .
\end{align}
Again only $\sqrt{2}\sigma_+ = \sigma+\tilde{\sigma}$ obtains a VEV. Defining $\tan\nu = \frac{m}{\mu_2}$,
there is a massless mode, 
\begin{equation}
\label{saturday}
\sigma_\parallel =
(\cos\nu \,\Xi-\sin\nu  \,{\sigma}_{+})\, ,
\end{equation} 
corresponding to the flat-direction that preserves $F_X=0$.
This is the adjoint degree of freedom whose fermionic superpartner will ultimately marry into the light 
Dirac state.  The $F$-term contribution to the potential gives a mass to the orthogonal mode,
\begin{equation}
 {\sigma}_\perp=
 ( \cos\nu\, {\sigma}_{+}+ \sin\nu \,\Xi),
\end{equation}
given by 
\begin{equation}
m_{\sigma_\perp}^{2}=(\mu_{2}^{2}+m^2),
\end{equation}
for the properly normalized adjoint field. The $D$-terms give a mass to $\Re(\sigma_{-})$ of 
$
m_{\Re(\sigma_{-})}=M_B $.
The $\Im(\sigma_{-})$ are massless Goldstone modes that
are eaten by the $B_{2\, \mu}$ gauge bosons (of $\sunnc_{F}\times\sunnc_{c}/\,\sunnc_{\sigma}$).
Writing $\sigma = \mu_2 \delta + \sqrt{2} \sigma^a T^a, \tilde{\sigma} = \mu_2 \delta_{ij} + \sqrt{2} \tilde{\sigma}^a T^a$ (where $T^{a}$ are
the generators of $\sunnc_\sigma$), $\sigma^{a}_\pm$ also then transforms correctly as an adjoint of the unbroken $\sunnc_{\sigma}$.
Finally the mass-squareds of the $\rho_{\pm}$ are found to be \begin{equation}
m_{\rho_{\pm}}^{2}=(\mu_{2}^{2}\mp\mu_{3}^{2}).\end{equation}
As we anticipated earlier, $\mu_{2}^{2}<\mu_{3}^{2}$ leads to a tachyon
(because the supersymmetry breaking is reduced by swapping the role
of some elements of $\sigma$ and $\rho$), while the choice $\mu_{2}^{2}=\mu_{3}^{2}$
has an enhanced global flavour symmetry and gives extra massless Goldstone
modes. 

Before continuing, we should point out that generically the $m \Xi X$ mass term is dangerous for ISS metastability: by acquiring VEVs on the diagonal, the $\Xi$ field is able to compensate for some of the $\mu_2^2$ couplings and render the rank condition inoperative.  However if $\sigma_\parallel$ has a 
local minimum at $X=\Xi=0$ (which we have eventually to check), then as long as the mass $m$ is small enough there is no danger. The onset of instability is where the $\rho_-$ becomes tachyonic, i.e. where $\Xi < 0$ with 
$|\Xi | > (\mu_2^2-\mu_3^2)/m $.  Hence \begin{equation} m\ll \mu_2 \label{flip} \end{equation} is required and will be assumed from now on.

Note that there is also of course the second breaking associated with the link-fields. 
This proceeds exactly as in the 
toy model, and it is not hard to show that the final light $\SU{n}$ is given by 
\begin{equation}
A_{\SU{n}}= \frac{1}{\sqrt{(g_c g_F)^2+(g_1 g_F)^2+(g_c g_1)^2}}\left( 
(g_c g_F) A_1 +(g_1 g_F) A_c+(g_c g_1) A_F \right) \, .
\end{equation}

Having identified the boson of the remaining unbroken gauge group, let us now turn to the fermion masses. 
There are 10 adjoint fermions of this final $\SU{n}$ in the model now, most of which will be getting supersymmetric masses from the various higgsings. We denote them as follows:
\begin{eqnarray}
\varphi_{11}\rightarrow \psi, & q_1,\tilde{q}_1 \rightarrow \eta_\pm, & \Xi \rightarrow \xi, \nn \\
X\rightarrow \chi, & \sigma_\pm \rightarrow \varsigma_\pm, & \lambda_{1,c,F} .
\end{eqnarray} 
The fermion masses from the first stage of breaking arise through 
 \begin{align}
\C{L} \supset& - \sqrt{2}\mu_2 (\chi \varsigma_+) -  g_F \sqrt{2} \tr (\lambda_F \varsigma \sigma^*) +  g_F \sqrt{2} \tr ( \tilde{\varsigma} \lambda_F  \tilde{\sigma}^*) + g_c \sqrt{2} \tr (\varsigma \lambda_c  \sigma^*) - g_c \sqrt{2} \tr (  \lambda_c \tilde{\varsigma} \tilde{\sigma}^*) \nn\\
\supset& - \sqrt{2}\mu_2 (\chi \varsigma_+) -  M_B \lambda_{B_2}^a \varsigma_-^a\, .
\end{align}
Including the contribution from the $\SU{n}_1$ breaking as well, we finally get the supersymmetric set of masses, 
\begin{align}
\C{L}  
\supset &  - m_{\sigma_\perp} (\chi^a \varsigma^a_\perp  ) - \mu_1 (\psi^a \eta_+^a) - M_A (\lambda_{B_1}^a \eta_-^a) - M_B (\lambda_{B_2}^a \varsigma _-^a)\, ,
\end{align}
where obviously $\varsigma^a_\perp =  
 ( \cos\nu\, {\varsigma}^a_{+}+ \sin\nu \,\xi^a)$ and where, to summarize,
\begin{eqnarray}
\lambda_{B_1} &=& \cos\vartheta \lambda_1 -\sin\vartheta \lambda_c \nn\\
\lambda_{B_2} &=& \cos\vartheta' \lambda_F -\sin\vartheta' \lambda_c \nn\\
M_A &=& \sqrt{2 (g_1^2+g_c^2)}\mu_1 \nn\\
M_B &=& \sqrt{2 (g_F^2+g_c^2)}\mu_2 \nn\\
\tan\vartheta=g_c/g_1 ; && \hspace{-0.6cm}\tan\vartheta'=g_c/g_F; \,\,\,\,\,  \tan\nu=m/\mu_2
\, .
\end{eqnarray}
At this stage there remains a massless pair of states which will be our Dirac gaugino (where we suppress the adjoint $a$-index); 
\begin{eqnarray}
\label{light-states}
\lambda_{\SU{n}} &\equiv & \frac{1}{\sqrt{(g_c g_F)^2+(g_1 g_F)^2+(g_c g_1)^2}}\left( 
(g_c g_F) \lambda_1 +(g_1 g_F) \lambda_c+(g_c g_1) \lambda_F \right) \nn\\
\bar{\lambda}_{\SU{n}} &\equiv & \varsigma_\parallel\, = 
(\cos\nu \, \xi- \sin\nu\,  \varsigma_+) \, .
\end{eqnarray}
The first of these is the superpartner of the light gauge boson, the second the superpartner  
of the flat direction in Eq.\eqref{saturday}.

Finally we are ready to consider the supersymmetry breaking contribution to the lightest pair of states. Clearly from Eq.\eqref{light-states} this will require a coupling between the $\lambda_c$ gauginos and $\varsigma_\parallel$. The relevant piece in the superpotential
that can generate this is 
\begin{eqnarray}
W^{(mag)}&\supset & 
\mu_{2}(Y\tilde{\rho}+\rho\tilde{Y})+ \sigma_+^a  
 \left(  \tilde{\rho} T^a  Y+ \tilde{Y}T^a \rho\right) 
 + Z \tilde{\rho}\rho 
\nn\\ &\equiv &
 \mu_{2}(Y\tilde{\rho}+\rho\tilde{Y})-\sin\,\nu  \, 
  \sigma_{\parallel}^a    \left(  \tilde{\rho} T^a  Y+ \tilde{Y}T^a \rho\right) 
 + Z \tilde{\rho}\rho 
 \, .\,\,\,\,
\label{eq:wfin2}
\end{eqnarray}
The model indeed has the required terms to generate a $m_D (\lambda^a_c\varsigma^a_\parallel)$ term but, as can be seen from Eq.\eqref{mdirac}, it is equivalent to a model with $h_1=h_2=1$ and so the contributions cancel. 

\begin{figure}[ht]
\begin{centering}
  \begin{picture}(330,100) (36,0)
    \SetWidth{2.5}
    \Text(95,57)[lb]{$f$}
    \Text(65,7)[lb]{$f$}
    \Text(132,7)[lb]{$\tilde{f}$}
    \Text(40,30)[lb]{$\tilde{f}$}
    \Text(152,32)[lb]{$\tilde{f}$}
    \Text(295,57)[lb]{$\tilde{f}$}
    \Text(265,7)[lb]{$\tilde{f}$}
    \Text(332,7)[lb]{${f}$}
    \Text(240,30)[lb]{${f}$}
    \Text(352,32)[lb]{${f}$}
    \Text(60,55)[lb]{$\kappa F$}
    \Text(120,55)[lb]{$\kappa^\dagger F$}
    \Text(255,55)[lb]{$\kappa^\dagger F$}
    \Text(320,55)[lb]{$\kappa F$}
    \Text(10,10)[lb]{$\Sigma$}
    \Text(30,-10)[lb]{$h$}
    \Text(230,-10)[lb]{$h^t$}
    \Text(150,-10)[lb]{$i\sqrt{2}g$}
    \Text(345,-10)[lb]{$-i\sqrt{2}g$}
    \Text(210,10)[lb]{$\Sigma$}
    \Text(180,10)[lb]{$\lambda$}
    \Text(380,10)[lb]{$\lambda$}
\includegraphics[angle=0,scale=.65]{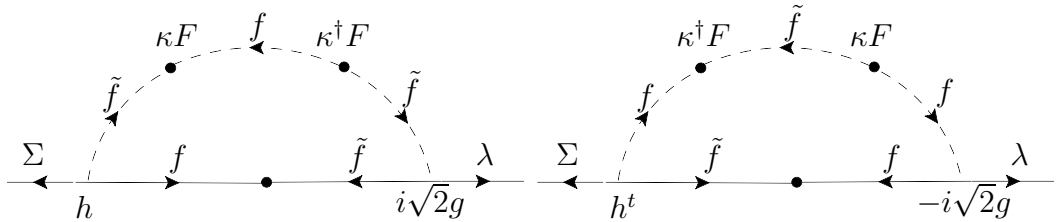}
\includegraphics[angle=0,scale=.65]{dirac-diag1.eps}
\end{picture}
\par\end{centering}
\caption{\it Diagrams contributing to Dirac mass terms at leading order in $F/M^2$.} 
\label{fig:dirac-diagrams}
\end{figure}

At this point we therefore make our final crucial modification: we assume that the Yukawa couplings also respect only the gauged part of the flavour symmetry, and are not the same for e.g. $Y$ and $\tilde{Y}$; 
\begin{eqnarray}
W^{(mag)}&\supset & 
 \mu_{2}( h_1 Y\tilde{\rho}+ h_2 \rho\tilde{Y}) -  
  \sin\nu \, \sigma_{\parallel}^a    \left(  h_1 \tilde{\rho} T^a  Y+ h_2 \tilde{Y}T^a \rho\right) 
 +  Z \tilde{\rho}\rho 
 \, .\,\,\,\,
\label{eq:wfin3}
\end{eqnarray}
Perhaps surprisingly, even though the couplings and the masses are multiplied by the same $h_i$ factor, the Dirac gaugino mass turns out to be non-zero: computing the diagrams in Figure~\ref{fig:dirac-diagrams} (or using the expressions from \cite{Benakli:2008pg,Benakli:2009mk}), one finds
\begin{equation}
m_D = \frac{\sqrt{2}}{64 \pi^2 } \frac{g_c \sin\nu \, F^2}{\mu_2^3}  \left| j\mbox{\small $(h^2_1,h^2_2)$}-j\mbox{\small $(h^2_2,h^2_1)$}\right|
\end{equation}
where 
\begin{equation}
j(a,b)=    \frac{a}{b(a-b)^3}\left( a^2-b^2 + 2ab \log\left| b/a \right| \right) \, . 
\end{equation}
In the limit as $h_1 \rightarrow h_2$ the approximate expression is 
\begin{equation}
m_D \approx  {g_c\sin\nu\, |h_1-h_2| } \frac{1}{16 \pi^2 } \frac{ F^2}{6\sqrt{2} \, .(h_2\mu_2)^3}. 
\end{equation}

\clearpage

We should now consider the scalar $\sigma_\parallel$ which could be tachyonic. Indeed the Coleman-Weinberg potential gives the $F$-flat direction a tachyonic mass-squared of order 
\begin{equation}
m^2_{\sigma_\parallel, CW} = - \frac{1}{16\pi^2} \sin^2\nu \,  \frac{\mu_3^4 }{\mu_2^2}\,. 
\end{equation}
However, assuming gauge couplings of order unity for the flavour groups, 
the terms that are induced in the K\"ahler potential in Eq.\eqref{mkahler} give a positive contribution; 
\begin{equation}
\label{mkahler2}
m^2_{\sigma_\parallel, K} \sim  \frac{1}{16\pi^2} \cos^2\nu \,\frac{\mu_3^4 }{\Lambda^2}\, .
\end{equation}
The latter contribution is dominant for 
$
\tan\nu \, = \frac{m}{\mu_2} <  \frac{\mu_2}{\Lambda},
$
or
\begin{equation}
{m} <  \frac{\mu^2_2}{\Lambda}\, .
\end{equation}
Note that this constraint automatically means that Eq.\eqref{flip} is satisfied; i.e. not only are there no tachyons, but the values of $\Xi$ where the $\rho_-$ becomes tachyonic are far away in field space. 

It is clear in the limit why this mechanism is bound to work. As we take $m\rightarrow 0$ the flat
direction is all $\Xi$ and the orthogonal massive direction is all $\sigma_+$. But only $\sigma_+$ is in contact with the supersymmetry breaking, and can get Coleman-Weinberg tachyonic mass-squared contributions, while the positive K\"ahler mass-squared contributions are all for the $\Xi$ direction.

In summary therefore, these three modifications, (i.e. a weakly coupled elementary adjoint meson in the electric theory, induced higher order K\"ahler potential terms of natural size, and an explicit breaking of the ungauged flavour symmetries in the couplings), give a pure Dirac gaugino whose mass is of order (assuming $\tan\nu = m/\mu_2 \ll 1$), \begin{equation}
m_D \sim \frac{1}{16 \pi^2 } \frac{g_c F^2}{\mu_2^3} \frac{m }{\mu_2} 
\, , \end{equation}
and non-tachyonic scalars.

\vspace{0.5cm}

\subsection{Scales}

Let us make some general observations about scales, and also address
the possibility of pseudo-Dirac masses. First of all, 
we have seen that the flat direction associated with the light gaugino mass is 
tachyonic unless we take $m/\mu_2 \ll 1$, and this gives a suppression 
of the Dirac gaugino mass by this same factor so that:
\begin{equation}
m_\lambda\approx m_{D}\sim\frac{ g}{16\pi^{2}}\left(\frac{\mu_{3}}{\mu_{2}}\right)^{4}m.
\end{equation}
A phenomenological requirement is that this mass should be similar
to or larger than the scalar masses. The beauty of this set up is that these are automatically screened as 
\begin{equation}
m_{\tilde{f}}^{2-loop}=\frac{g^{2}}{16\pi^{2}}\frac{|F_{Z}|}{M}\frac{\sqrt{\left\langle q_{1}\tilde{q}_{1}\right\rangle }}{M}=\frac{g^{2}}{16\pi^{2}}\left(\frac{\mu_{3}}{\mu_{2}}\right)^{2} \mu_{1}.\end{equation}
Assuming that $g\sim$1, then for these to be the same
we require $\Lambda\gg \mu_{2}\gtrsim\mu_{3}\gg m \gtrsim \mu_{1}$ with 
\begin{eqnarray}
m &\sim &  \left(\frac{\mu_{2}}{\mu_{3}}\right)^{2} \mu_1\nn \\
m & \lesssim & \frac{\mu_2^2 }{\Lambda}\, .\label{eq:murel}
\end{eqnarray}
In the limit $\mu_2\gtrsim \mu_3 $ and $\mu_1/m \rightarrow 0$ we get dominant gaugino masses. Thus 
the screening effect of gaugino mediation can counteract the 
fact that the gaugino masses are suppressed. Note that neither the gaugino nor the scalar mass depends on the absolute values of $\mu_2$ and $\mu_3$: they only depend on the ratio.

As an example of scales, suppose we allow a
hierachy of order 10 so that $\mu_2 / \mu_3 \sim 10$.
Then since phenomenology requires $M_{Dirac}\sim1$TeV we have $m\sim10^{6}$TeV.
The scalar masses are similar if $\mu_1 \sim 10^4$TeV. There is then a wide variety of suitable values for $\mu_3$, $\mu_2$
and $\Lambda$: for example $\mu_3=10^7$TeV, $\mu_2=10^8$TeV and $\Lambda=10^{10}$TeV satisfy all the constraints. 
Note that the only phenomenologically required tuning is the usual one, of getting $M_{Dirac}\sim 1$TeV. The other parameters can fall within wide ranges. 

Similar gaugino and scalar masses are perhaps more natural if there is an underlying reason for having $\mu_2 \sim \mu_3\gg\mu_1 \sim m$. In this case $m\sim \mu_1\sim 100$TeV gives masses of order 1TeV for gauginos and scalars. Avoiding tachyons requires $\Lambda < \frac{\mu_2^2}{m}$ which is very easy to satisfy. For example taking $\mu_2\sim\mu_3\sim  10^5$TeV, the no-tachyon constraint becomes only that $\Lambda < 10^8$TeV.

On the other hand, when the two-loop sfermion masses are very screened, then the sfermion masses can become dominated by the three-loop supersoft masses:
\begin{align}
m_{\tilde{f}}^{3-loop}=& \frac{g m_D}{2\pi} \sqrt{\log [\frac{m_R^2}{m_D^2}]} \nn\\
\sim& \frac{\alpha}{8\pi^2} \left( \frac{\mu_3}{\mu_2}\right)^4 m \sqrt{ \log [ \frac{4\pi \alpha \mu_2^8}{m^2 \Lambda^2 \mu_3^4}]}.
\end{align}
For the examples above the logarithmic term contributes an $\C{O}(1)$ factor, giving \begin{equation}
m_{\tilde{f}}^{3-loop}\sim 0.1 m_D \sim 10^2{\rm GeV} \end{equation}
independently of the screening.

\subsubsection{Pseudo-Dirac Gauginos}

Now let us comment on the possibility of pseudo-Dirac gauginos. Thus far the model possesses
a residual $R$-symmetry which is a combination of the exact $R$-symmetry
of Table \ref{blocked} and an anomalous $\U1_{A}$ symmetry and so
Majorana mass-terms  will not arise. Bearing in mind the discussion of
Section \ref{sec:R-emergency}, a simple and natural way to have a
Majorana component is to invoke an emergent $R$-symmetry by adding
an additional pair of messenger fields $f_{3}$, $\tilde{f}_{3}$
that are charged only under the second flavour $\sunnc_{F}$. These
can couple as in Section~\ref{sec:R-emergency} so as to generate a small additional
term in the superpotential, \begin{equation}
W^{(mag)}\supset\lambda'\, f_{3}\tilde{f}_{3}Z+M_{f_{3}}f_{3}\tilde{f}_{3}\, , \end{equation}
without disrupting the supersymmetry breaking, and will generate a
Majorana mass term of order \begin{equation}
M_{Majorana}=\frac{\lambda'g^{2}}{16\pi^{2}}\frac{\mu_{3}^{2}}{M_{f_{3}}}.\end{equation}
Note that in the absence of messenger couplings to the adjoint there can be no
Majorana mass for $\Sigma$, so that the gaugino mass matrix takes
the characteristic see-saw form, \begin{equation}
M_{gaugino}\sim\left(\begin{array}{cc}
M_{Majorana} & m_{D}\\
m_{D} & 0\end{array}\right),\end{equation}
familiar from neutrino physics. Indeed if one were to allow $m_{D}\ll M_{Majorana}$
a see-saw suppression of gaugino masses would ensue. Given the general
arguments of Section \ref{sec:R-emergency}, one might think this
would be unnatural without a source of spontanous $R$-breaking  because the $R$-breaking contribution
would be dominant, however this is not necessarily the case.
Indeed the Majorana mass is given by \begin{equation}
M_{Majorana}\sim\frac{\lambda'g^{2}}{16\pi^{2}}\left(\frac{\mu_{3}}{M_{f_{3}}}\right)\mu_{3}.\end{equation}
The Majorana mass can be comparable to or larger than the Dirac mass-term
when \begin{equation}
M_{f_{3}}\lesssim\lambda'\left(\frac{\mu_{2}}{\mu_{3}}\right)^{4}
\frac{\mu_3^2}{m}.\end{equation}
Note that $\lambda'$ is the parameter associated with the breaking
of $R$-symmetry so it is naturally expected to be small and for much
of the parameter space the gaugino is indeed pseudo-Dirac. However
this can be off-set in the regions where 
$\mu_{3}\gg m$. (Recall that the absolute value of $\mu_3$ does not affect the masses.) 
For example, for our previous example of $\mu_2\gtrsim \mu_3 \sim 10^5$TeV and $\mu_1\sim m \sim 100$TeV, 
the Majorana mass-terms are comparable to the Dirac ones when $M_{f_{3}}\lesssim\lambda'\,10^{8}$TeV.
In order for the $f_{3}$, $\tilde{f}_{3}$ messengers to evade detection
we impose $M_{f_{3}}>1$TeV, which allows Majorana masses to be comparable
even for $R$-symmetry breaking as small as $\lambda'\sim10^{-8}$. 

In general therefore there is no naturalness argument that prevents
gauginos still having predominantly Majorana masses rather than pseudo-Dirac
ones. From an $R$-symmetry perspective it is easy to see why: in
building the model we broke the exact $R$-symmetry \emph{twice} --
once to generate the metastable supersymmetry breaking (via the tiny
electric-quark mass terms which led to the $\mu_{ISS}^{2}$ operators
of the magnetic theory), and once to generate the Majorana mass term
(via the $\lambda'$ operator). There is no indication as to which
effect should be dominant. 

There is however an interesting lower bound on the degree of {}``pseudo-ness''
in the pseudo-Dirac masses due to Majorana contributions coming from
anomaly mediation, and this leads to a definite prediction of pseudo-Dirac
masses in the case that the theory in the absence of gravity preserves
an exact $R$-symmetry.\footnote{Note that this requires this requires an R-symmetric Higgs sector such as the MRSSM \cite{Kribs:2007ac}. }
Indeed it tells us that exact Dirac masses
are not possible, and given certain common assumptions gives a precise and measurable prediction of
the degree of splitting in the mass spectrum of pseudo-Dirac gaugino
pairs. Namely, if we assume that the supergravity K\"ahler potential for our matter fields is canonical, then we can take the potential to be 
\begin{align} 
V = |F|^2 + V_F^{hid} - 3 m_{3/2}^2 M_P^2 
\end{align}
where $F$ is the visible-sector $R$-symmetry preserving $F$-term $\sim \mu_3^2$ for our model above, $V_{F}^{hid}$ is the potential from the hidden sector where R-symmetry is broken and the gravitino mass $m_{3/2}$ is generated. As a minimum value for the splitting, we shall take $V_{F}^{hid} = 0$ and so assuming that the cosmological constant is (approximately) zero we must have $m_{3/2} = |F|/\sqrt{3}M_P$. Then since we are assuming canonical K\"ahler potentials, anomaly mediation will generate Majorana gaugino masses $m_{1/2}^i$ for group $i$ with beta-function coefficient $b^i$ given by
\beq
m_{1/2}^i = \frac{\alpha^i b^i}{4\pi}  m_{3/2} = \frac{\alpha^i b^i}{4\pi} \frac{\mu_3^2}{\sqrt{3} M_P}.
\eeq
Taking $\mu_3 = 10^6\, \mathrm{TeV}$ gives a gravitino mass of ${\cal O}(1$~GeV$)$ and a splitting of order an MeV. Clearly much smaller splittings are possible especially since, as noted above, only the ratio $\mu_3/\mu_2$ affects the spectrum. Such small splittings could have interesting consequences for dark matter or experimental signals at the LHC, but we leave this to future work. 

\subsection{$G_{vis} = \SU{3} \times \SU{2}\times \U1$ and bachelor states}
\label{sec:bachelors}

So far we have been working in the $\SU{5}$ GUT framework for simplicity. Some care is required in going to the 
Standard Model $G_{vis} = \SU{3} \times \SU{2}\times \U1$ gauge group, since in contrast to the case of Majorana gauginos the entire model has to be modified to accommodate the change.
 
To see why, let us elaborate on a subtlety when the flavour gauge groups are not identical to the colour group.  A potential problem arises due to the presence of bifundamental states charged under different groups, specifically in $(\bf{3}, \bf{\bar{2}})$ and $(\bf{\bar{3}},\bf{2})$ representations: these are termed ``bachelor'' states, since they also arise from the decomposition of a $\mathbf{24}$ of $\SU{5}$ to $\bf{8}_0 + \bf{3}_0 + \bf{1}_0 + (\bf{3},\bf{2})_{-5/6} + (\bf{\bar{3}},\bf{2})_{5/6}$. Consider the toy model of section \ref{sec:generalmodel} but where we take $G_{vis}$ to be $\SU{3} \times \SU{2}\times \U1$ and $G_{hid}$ to be $\SU{5}$. We can decompose the gauge and link-fields into matrix form as  
\beq
A^{hid} = \begin{pmat}({..|..})
  & & &  & \cr
  & A_{3\bar{3}}^{hid} & & A_{3\bar{2}}^{hid} & \cr
  & & &  & \cr\-
  & A_{3\bar{2}}^{2,\dagger} & & A_{2\bar{2}}^{hid}  & \cr
  & & &  & \cr
\end{pmat}
,\,
A^{vis} = \begin{pmat}({..|..})
  & & &  & \cr
  & A_{3\bar{3}}^{vis} & & 0 & \cr
  & & &  & \cr\-
  & 0 & & A_{2\bar{2}}^{vis}  & \cr
  & & &  & \cr
\end{pmat}
,\,  
L,\tilde{L} = \begin{pmat}({..|..})
  & & &  & \cr
  & L_{3\bar{3}},\tilde{L}_{3\bar{3}} & & L_{3\bar{2}},\tilde{L}_{3\bar{2}} & \cr
  & & &  & \cr\-
  & L_{2\bar{3}},\tilde{L}_{2\bar{3}} & & L_{2\bar{2}},\tilde{L}_{2\bar{2}}  & \cr
  & & &  & \cr
\end{pmat}.
\eeq
Note that as before in our definition of $L$ the first and second indices are the first and second gauge groups respectively, while for $\tilde{L}$ they are the second and first groups. We trust this will confuse the reader. Then higgsing with a VEV for $L, \tilde{L}$ of $\mu_\ell$, the bifundamental supersymmetric fermion mass terms -- explicitly showing the representations as indices -- are:
\begin{align}
\C{L} \supset& - \mu_\ell \xi_{2\bar{3}}\tilde{\eta}_{3\bar{2}}  + \mu_\ell\eta_{2\bar{3}}\xi_{3\bar{2}}- m \varsigma_{3\bar{2}} \xi_{2\bar{3}} + g_2 \mu_\ell \tr(\eta_{2\bar{3}} \lambda^2_{3\bar{2}}) - g_2 \mu_\ell \tr( \lambda^2_{3\bar{2}} \tilde{\eta}_{2\bar{3}}) + (2\leftrightarrow 3) \nn\\
\supset& - (\sqrt{2} \mu_\ell\eta_{3\bar{2}}^+ + m  \varsigma_{3\bar{2}})\xi_{2\bar{3}}   + \sqrt{2}g_2 \mu_\ell\eta_{2\bar{3}}^- \lambda^2_{3\bar{2}} + (2\leftrightarrow 3)\, .
\end{align}
The reader can easily see that the gaugino components are made massive at the higgsing scale, and leave two massless adjoint fermions. The problem is easily avoided in the toy model by always taking $G_{vis}$ and $G_{hid}$ to be be identical, as we stated earlier.

The satisfactory configuration for the UV complete model of subsection \ref{subsec:model} can be slightly different however. We actually have several options, which all require a gauge group $G_{vis}^\prime$ in the quiver diagram, either to replace one of the flavour or colour groups or to supplement them. 
The most straightforward possibility is to perform the dualities separately for $\SU{2}$ and $\SU{3}$ groups, with separate supersymmetry breaking, and have no link-fields transforming under both. Equivalently one can use a form of Seiberg duality  that admits a spontaneous breaking of the GUT symmetry of the colour group down to $G_{vis}^\prime$. The duality of \cite{K:Adj,KS:DKSS,KSS:DKSS} (KSS) allows one to do this in a simple way. In this form of duality one introduces an elementary 
$\mathbf{24}$ of $\SU{n}_c$ and writes a cubic superpotential for it that generates the usual VEV along the diagonal (we should add that this entails an explicit but controlled breaking of $R$-symmetry -- but it should not effect the rest of the theory). Both the electric and magnetic theories are then higgsed down to a product group, with the magnetic colour group being 
$\SU{3} \times \SU{2}\times \U1$  and the electric dual group
being $\SU{N_f-3}\times  \SU{N_f-2}\times \U1$. The offending off-diagonal components of the link-fields are all made heavy by the GUT higgsing. (There are also additional mesons in this theory, but of course they can be all made massive by Yukawa couplings in the electric theory, in exactly the same way as in canonical Seiberg duality.) An advantage of using KSS duality is that the $\SU{\tilde{N}}$ flavour node can be empty  in the free-magnetic window. In addition, in the limit where $\alpha_{vis} \gg \alpha_{hid}$ the unification observed in the SM sector is purely dual-unification as described in Section~\ref{unif-discussion}.

Another possibility is to take the colour group as $\SU{5}$ and replace $\SU{n}_F$ with $\SU{3} \times \SU{2}\times \U1$ (although this leads to problems for the $\U1$ adjoint scalars whose masses are not lifted by the additional K\"ahler terms). However there are many alternatives: we outline the most straightforward below.

\subsubsection{The preferred model:  $\SU{N_f} \rightarrow G_{vis} \times G_{vis}^\prime \times \SU{n}_F \times \SU{\tilde{N}}$}
\label{subsec:finalmodel}

The model above in \ref{subsec:model} required $\tilde{N} \ge 6$ for the colour group to be infra-red free. This then leaves many light states in the hidden, supersymmetry breaking sector. However, by \emph{introducing another flavour node} $G_{vis}^\prime$ that is a copy of the MSSM gauge groups we can at once cure the bachelor problem above and allow $\tilde{N} \ge 1$. 

\begin{figure}[htp]
\begin{center}
\epsfig{file=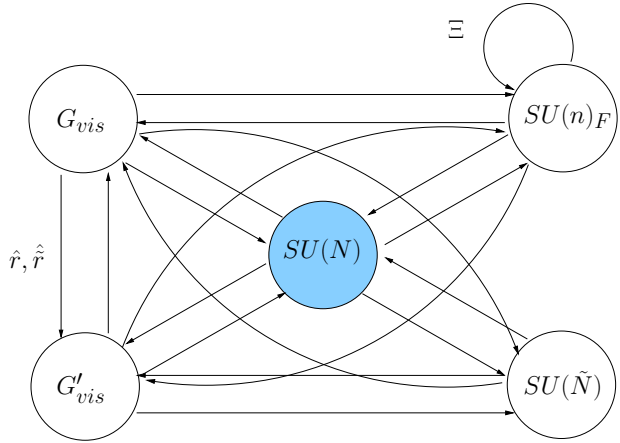,width=0.5\linewidth}
\parbox{13cm}{
\caption{\em \label{compuv2}The UV completion of the model with $G_{vis} = G_{vis}^\prime = SU(3)\times SU(2)\times U(1)$. }
}
\end{center}
\end{figure}

We must specify the elementary mesons in the electric theory that are integrated out along with the link-fields between flavour groups. While the choice of these is obvious for those linking $\SU{n}_F$ with $\SU{n}_1$, we must specify that in addition to those bifundamental under all the $G_{vis}$ groups and the additional $\SU{\tilde{N}}$ left after higgsing $\SU{\tilde{N}}\times \SU{n}_F$ required for supersymmetry breaking, there are only fields in representations of $\hat{r}_{3\bar{3}} \rightarrow (\mathbf{3}, \mathbf{\ov{3}})_{-1/3,1/3}, \hat{\tilde{r}}_{3\bar{3}} \rightarrow (\mathbf{\ov{3}}, \mathbf{3})_{1/3,-1/3}, \hat{r}_{2\bar{2}}\rightarrow(\mathbf{2}, \mathbf{\ov{2}})_{1/2,-1/2},\hat{\tilde{r}}_{2\bar{2}}\rightarrow(\mathbf{\ov{2}}, \mathbf{2})_{-1/2,1/2}$. To be concrete, the electric superpotential is given by
\beq
W^{(elec)} = m_I^J Q^I \tilde{Q}_J + S_I^J Q^I\tilde{Q}_J  
\eeq
where the mesons $S_i^a$ transform under $\SU{3}\times\SU{2}\times \SU{3}^\prime\times\SU{2}^\prime \times \SU{5}_F \times \SU{\tilde{N}}$ as
\beq
m_I^J= \frac{1}{\Lambda}\begin{pmat}({|||||})
\mu_1^2 & &   &  &  &\cr\-
& \mu_1^2 & & & &\cr\-
& & \mu_4^2& & & \cr\-
 && & \mu_4^2&& \cr\-
 && & & \mu_2^2 & \cr\-
 && & && \mu_3^2\cr
\end{pmat},\qquad S_I^J = \begin{pmat}({|||||})
 & &\hat{r}_{3\bar{3}}   &  & S_{3\bar{5}} &S_{3\bar{\tilde{N}}}\cr\-
&  & &\hat{r}_{2\bar{2}} & S_{2\bar{5}} &S_{2\bar{\tilde{N}}}\cr\-
\hat{\tilde{r}}_{3\bar{3}} & & & & S_{3'\bar{5}} &S_{3'\bar{\tilde{N}}} \cr\-
 & \hat{\tilde{r}}_{2\bar{2}}& & &S_{2'\bar{5}} &S_{2'\bar{\tilde{N}}} \cr\-
\tilde{S}_{5\bar{3}}&  \tilde{S}_{5\bar{2}}& S_{5\bar{3}'} &S_{5\bar{2}'} & \Xi & \cr\-
\tilde{S}_{\bar{\tilde{N}}3}& \tilde{S}_{\bar{\tilde{N}}2}& S_{\tilde{N}\bar{3}'}&S_{\tilde{N}\bar{2}'} & & \cr
\end{pmat}.
\eeq
The quiver diagram is shown in figure \ref{compuv2}. 
%\clearpage 

\begin{figure}[htp]
\begin{center}
\epsfig{file=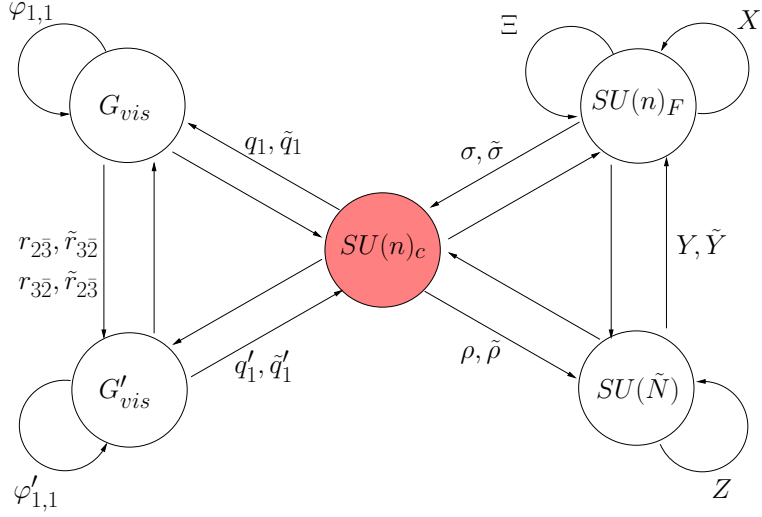,width=0.6\linewidth}
\parbox{13cm}{
\caption{\em \label{compir2} The magnetic Seiberg dual of the quiver in Figure~\ref{compuv2}. Supersymmetry is broken by the rank-condition with the $\SU{\tilde{N} }$ node providing the $F$-term breaking: $F_Z\neq 0$. }
}
\end{center}
\end{figure}

%\clearpage 

In the magnetic dual theory, the elementary mesons (except for $\Xi$) are integrated out along with their partner mesons, leaving
massless magnetic mesons $r_{2\bar{3}}, \tilde{r}_{3\bar{2}}, r_{3\bar{2}}, \tilde{r}_{2\bar{3}}$ with superpotential terms
\begin{align}
W^{(mag)} \supset& r_{2\bar{3}} (a_1 q_{1, 3\bar{3}} q^\prime_{1, 3 \bar{2}} + b_1 q_{1, 3\bar{2} } q^\prime_{1,2 \bar{2}}) + ( a_2 \tilde{q}_{1, 3\bar{2}}\tilde{q}^\prime_{1,2\bar{2}}  +  b_2\tilde{q}_{1, 3\bar{3}}\tilde{q}^\prime_{1,3\bar{2}} )\tilde{r}_{2\bar{3}} + ( 2 \leftrightarrow 3)
\end{align}
where we have included different $\C{O}(1)$ coefficients $a_i, b_i$. The quiver diagram is shown in figure \ref{compir2}, with the fields labelled. After the higgsing of the colour group and $\SU{5}_F$ at the scale $\mu_2$ to a diagonal $\SU{5}_\sigma$ with a Dirac gaugino mass for the lightest gaugino state, we then higgs this group with $G_{vis}^\prime$ at the scale $\mu_4$ and the remaining group with $G_{vis}$ at the scale $\mu_1$. The relevant fermionic mass terms are
\begin{align}
\C{L} \supset& - \sqrt{2} \mu_1 q_{1,3\bar{2}}^+ \psi_{2\bar{3}}  - \sqrt{2} \mu_4 (q_{1,3\bar{2}}^\prime)^+ \psi^\prime_{2\bar{3}} \nn\\
& - m_D \varsigma_{3\bar{2}} \lambda_{2\bar{3}}    + \sqrt{2}g_\sigma \mu_1 q_{1, 3\bar{2}}^- \lambda_{2\bar{3}} + \sqrt{2} g_\sigma \mu_4 (q_{1,3\bar{2}}^\prime)^- \lambda_{2\bar{3}} \nn\\
& - \mu_1 (a_1 r_{2\bar{3}}q_{1,3\bar{2}}^\prime + b_1 \tilde{q}_{1,3\bar{2}}^\prime\tilde{r}_{2\bar{3} } ) - \mu_4 (a_2 r_{2\bar{3}} q_{1, 3\bar{2} }   + b_2\tilde{r}_{2\bar{3}} \tilde{q}_{1, 3\bar{2}} ) \nn\\
& + ( 2 \leftrightarrow 3) 
\end{align}
Here $\psi^\prime$ is the fermion of the superfield $\phi_{1,1}^\prime$, $\varsigma$ is now the linear combination of the fermions from $\varsigma_+$ and $\xi$ that gives our supersymmetry breaking Dirac mass, and $g_\sigma$ is the gauge coupling of $SU(n)_\sigma$.
The mass matrix is then non-degenerate  provided that $a_1 b_2 - a_2 b_1 \ne 0$. Assuming that $\mu_4 \gg \mu_1$  we have the masses for the (Dirac) gaugino bachelors of
\begin{align}
m_{\lambda}^{bachelor} \approx& \, m_D \frac{\mu_1^2}{g_\sigma \mu_4^2} \frac{(a_1 b_2 - a_2 b_1)}{2a_2 b_2}\nn\\
\approx& \, \frac{1}{16\pi^2} \frac{\mu_3^4}{\mu_2^4} \left(\frac{\mu_1}{\mu_4}\right)^2 m
\end{align}
but perhaps more naturally we can choose $\mu_4 = \mu_1$, giving 
\beq
m_{\lambda}^{bachelor} = m_D [ 1 + \C{O}(g^2, (m_D/\mu_1)^2)]. 
\eeq

This therefore provides an elegant solution to the bachelor mass problem while allowing the supersymmetry breaking sector to be extremely simple, with $\tilde{N} =1$ if desired. However, it is also possible to modify the supersymmetry breaking sector of the model without reintroducing the problem of bachelor masses: so long as the theory is higgsed to an $\SU{n}_\sigma$ gauge group with Dirac gaugino masses at a scale above $\mu_1$ and $\mu_4$ then all of the above holds.

\section{Conclusions}

A simple model of Dirac gauginos was presented based on a two site
``deconstructed gaugino mediation'' model.
 The model preserves an $R$-symmetry, thereby evading the metastability
issue that is directly linked to the generation of Majorana masses for
gauginos. A UV completion was also presented
by adapting ISS metastable supersymmetry breaking. This results in a
comprehensive model that, as well as the supersymmetry breaking, generates
the necessary additional adjoint degrees of freedom as quarks of the
magnetic Seiberg dual ISS theory.
Further, the ISS framework predicts higher order operators in the K\"ahler
potential that are able to prevent the appearance of the problematic
tachyons typically occuring in Dirac gaugino models (along the flat
directions corresponding to the superpartners of the new light fermionic
adjoints).

The spectrum has an unusual lack of dependence on the magnitude of
supersymmetry breaking due to a ``screening'' that can take place for both
the gauginos and the scalars. For example, in the UV complete theory, the supersymmetry breaking
sector has a linear meson term split into 3 flavour blocks with parameters
$\mu^2_{i=1..3}$, and with the non-zero $F$-term being $\mu_3^2$. In terms
of these parameters the light scalar mass is
\begin{equation}
m_{\tilde{f}}\sim
\frac{g^{2}}{16\pi^{2}}\left(\frac{\mu_{3}}{\mu_{2}}\right)^{2}
\mu_{1} \, , \end{equation}
upto group theoretical factors, and
assuming that $\mu_1$ is chosen to be large enough that this two-loop contribution is still dominant over the three-loop ones. 
The Dirac gaugino mass is
\begin{equation}
m_{\lambda}\sim \frac{ g}{16\pi^{2}}\left(\frac{\mu_{3}}{\mu_{2}}\right)^{4}m,
\end{equation}
where $m$ is an arbitrary mass parameter related to a Yukawa coupling in the
UV completion. Importantly neither quantity depends on the absolute value of the supersymmetry breaking 
$\mu_3^2$, but just on the ratio $\mu_3/\mu_2$.

A controlled breaking of $R$-symmetry can be
introduced to make the gauginos arbitrarily pseudo-Dirac. Finally we also
pointed out that the $R$-symmetry breaking associated with the
cancellation of the cosmological constant is mediated to the
Standard-Model by anomaly mediation, providing a lower bound on how purely
Dirac the gauginos can be.

\subsection*{Acknowledgements}

MDG acknowledges support from the German Science Foundation (DFG) under SFB 676, and is grateful to the CERN Theory Division for hospitality. He would like to thank Karim Benakli for discussions and collaboration on related projects. SAA is grateful to the Leverhulme Trust for support. We thank Tony Gherghetta for permission to reproduce Figure 1. We are grateful to Zohar Komargodski for valuable discussions, and to the organisers of the Brussels workshop on Gauge mediation of supersymmetry breaking.

\end{document}